\documentclass[preprint,preprintnumbers,superscriptaddress,showkeys,showpacs,nofootinbib,aps,floatfix]{revtex4}

\usepackage{amssymb}
\usepackage{epsfig}
\usepackage{amssymb}
\usepackage{amsmath}
\usepackage{epsf}
\usepackage{epsfig}
\usepackage{color}
\usepackage{ifpdf}
\usepackage{graphics}
\usepackage{longtable}
\usepackage{multirow}

\newcommand{\nn}{\nonumber}
\newcommand{\ba}{\begin{eqnarray}}
\newcommand{\ea}{\end{eqnarray}}
\newcommand{\be}{\begin{equation}}
\newcommand{\ee}{\end{equation}}
\newcommand{\bd}{\begin{displaymath}}
\newcommand{\ed}{\end{displaymath}}

\newcommand{\old}[1]{}

\newcommand{\plotangle}{270}

\def\kek{High Energy Accelerator Research Organization (KEK), Tsukuba 305-0801, Japan}
\def\Sokendai{School of High Energy Accelerator Science, The Graduate University for Advanced Studies (Sokendai), Tsukuba 305-0801, Japan}

\def\kyoto{Yukawa Institute for Theoretical Physics, Kyoto University, Kyoto 606-8502, Japan}
\def\columbia{Physics Department, Columbia University, New York, NY 10027, USA}

\begin{document}

\preprint{CU-TP-1206}
\preprint{KEK-CP-316}
\preprint{YITP-14-96}

\title{Timelike pion form factor in lattice QCD}

\author{Xu~Feng}
\affiliation{\kek}
\affiliation{\columbia}

\author{Sinya~Aoki}
\affiliation{\kyoto}

\author{Shoji~Hashimoto}
\affiliation{\kek}
\affiliation{\Sokendai}

\author{Takashi~Kaneko}
\affiliation{\kek}
\affiliation{\Sokendai}

\begin{abstract}
  We perform a nonperturbative lattice calculation of the complex
  phase and modulus of the pion form factor in the timelike momentum 
  region using the finite-volume technique.
  We use two ensembles of $2+1$-flavor overlap fermion at pion
  masses $m_\pi$ = 380 and 290~MeV.
  By calculating the $I=1$ correlators in the center-of-mass and three 
  moving frames, we obtain the form factor at ten different values of
  the timelike momentum transfer around the vector resonance.
  We compare the results with the phenomenological model of
  Gounaris-Sakurai and its variant.
\end{abstract}


\maketitle

\section{Introduction}
Lattice quantum chromodynamics (QCD) has been successful at providing
first-principles calculations of various physical quantities, among
which the calculations of the so-called {\em gold-plated} quantities,
such as the lowest-lying hadron masses, decay constants and matrix
elements with one hadron or vacuum as the initial or final state, are
carried out with controlled errors.
On the other hand, there are many interesting physical observables
that are beyond {\em gold-plated}.
An interesting example is that of transition amplitudes involving
non-QCD initial/final states, such as the amplitudes for
$\eta_c,\chi_{c0}\rightarrow\gamma\gamma$ \cite{Dudek:2006ut} and
$\pi^0\rightarrow\gamma\gamma$
\cite{Cohen:2008ue,Shintani:2011vc,Feng:2011za,Feng:2012ck,Lin:2013im}. 
Another example is the $K\rightarrow\pi\pi$
decay~\cite{Blum:2011ng,Blum:2012uk,Boyle:2012ys}, where the final
state consists of multiple strongly interacting pions.
For such cases, the finite-volume correction to the two-body state
must be properly taken into account \cite{Lellouch:2000pv}.

For the $K\rightarrow\pi\pi$ decay, the main efforts have been made to
reproduce the physical amplitude where the center-of-mass (CM) energy
of the two pions, $E^*$, is equal to the kaon mass $m_K$. 
In this work, on the other hand, we study a simpler quantity, the
timelike pion form factor,
for which the final state contains two pions but its energy $E^*$ 
varies in the whole $\pi\pi$ elastic scattering region.

Physically, the timelike pion form factor describes how an
electromagnetic vector current couples to two pions. 
We concentrate on the isovector part of the electromagnetic current,
which associates with an isospin $I=1$ $\pi\pi$ scattering state.
The corresponding $\pi\pi$ scattering phase has been
studied by several lattice groups using different techniques
\cite{Aoki:2007rd,Gockeler:2008kc,Feng:2010es,Lang:2011mn,Aoki:2011yj,Pelissier:2012pi,Dudek:2012xn}.

Besides the tests of the lattice calculations of multiparticle
states, the pion form factor provides information on the 
electromagnetic structure of pions. 
At tree level, the coupling of an electromagnetic current to
spinless pointlike particles is completely determined by their charge.
For the composite particles such as the pion, however, one
must take into account their internal structure, which is described by a
form factor depending on the momentum transfer, the so-called
electromagnetic form factor. 
A direct lattice QCD calculation of the pion form factor can reveal
this internal structure of the pion.
Experimentally, the timelike pion form factor can be measured through
the process $e^+e^-\to\pi^+\pi^-$, and it shows a resonance structure
due to the $\rho$ meson.
It is therefore interesting to calculate the whole functional form on
the lattice and compare it with the available experimental data.

Previous lattice calculation of the pion form factor has been carried out
at Euclidean (or spacelike) momenta, $q^2<0$
\cite{Brommel:2006ww,Frezzotti:2008dr,Boyle:2008yd,Aoki:2009qn,Nguyen:2011ek,Brandt:2013dua,Koponen:2013boa}. 
At low momenta $q^2\rightarrow 0^-$ the pion charge radius can be extracted. 
In this work, we calculate the pion form factor in the timelike momentum region,
which provides a different approach to extract the charge radius from 
the opposite direction $q^2\rightarrow 0^+$. 

The method to calculate the amplitudes or the form factors involving
two particles in the final state was originally proposed by
\cite{Lellouch:2000pv} and extended to moving frames by
\cite{Christ:2005gi,Kim:2005gf}. 
All these works chose $K\rightarrow\pi\pi$ as the process to study,
where the initial state is an on-shell kaon and the final state
consists of $\pi\pi$ in the $I=0$ or $2$ channel. 
In \cite{Meyer:2011um}, it is proposed to extract the pion form factor
from the process $\gamma^*\rightarrow\pi\pi$, 
where the initial state is a virtual photon and the two pions form a
$P$-wave scattering final state in the $I=1$ channel. 
In this work we adopt this approach and extend it to the moving frames,
which allow us to obtain the form factor in the whole elastic
$\pi\pi$ scattering region.    

The methods described above and used in our calculation are universal
and can be applied to other physical observables involving
two-particle initial or final state. 
A direct extension is the timelike scalar form factor of the pion.
In this case, the interest is in the $I=0$ scalar channel, where the
sigma resonance is relevant.
If we consider two particles with unequal masses, the method may be
extended to the $K\pi$ system. 
The timelike form factor is then related to the process of
semileptonic $\tau$ decays $\tau\rightarrow K\pi\nu_\tau$, where a
weak current couples to $K\pi$ and a resonance $K^*$ appears in this
channel. 
One may also extend the calculation from the meson sector to the
baryon sector, such as the timelike nucleon form factor
associated with the process $e^+e^-\rightarrow p\bar{p}$.  

Since most of the hadrons, such as $\rho$, $K^*$ and $\Delta$, are
resonances, one should treat them as a multiparticle system in the
lattice calculation.
In this regard, our exploratory study of the timelike pion form
factor provides a test of the lattice method and helps to pave the way
towards more challenging calculations with full consideration of 
more complicated resonance physics.

This paper is organized as follows. 
In Sec.~\ref{sect:form_factor} we introduce some phenomenological
background of the timelike pion form factor. 
In Sec.~\ref{sect:FSM} we discuss the finite-volume method used in
our calculation. 
Then, in Sec.~\ref{sect:corr_func} we give the construction of the
interpolating operator and the correlation function. 
The analysis of lattice results is described in
Sec.~\ref{sect:analysis}.

\section{Timelike pion form factor}
\label{sect:form_factor}
Hadron production via virtual photon in $e^+e^-$ annihilation offers
a fundamental test of QCD.
At low energies, the dominant hadronic final state consists of two
charged pions. 
The total cross section $\sigma(e^+e^-\rightarrow\pi^+\pi^-)$ is given by
a square of the modulus of the electromagnetic pion form factor $F_\pi(s)$,
\begin{equation}
  \label{eq:not_point_like}
  \sigma(e^+e^-\rightarrow\pi^+\pi^-)
  =\sigma^0(e^+e^-\rightarrow\pi^+\pi^-)|F_\pi(s)|^2,
\end{equation}
where $\sigma^0(e^+e^-\rightarrow\pi^+\pi^-)$ is the tree-level cross
section calculated with scalar QED by assuming that the pion is a
pointlike particle. 
The QCD corrections are all encoded in the pion form factor
$F_\pi(s)$, which describes how a (virtual) photon couples to two
pions in the final state.

The pion form factor is defined by a vector matrix element between 
the QCD vacuum and the $\pi\pi$ in and out states
\begin{eqnarray}
  \label{eq:matrix_element}
  \langle\pi^+({\bf p}_+)\pi^-({\bf p}_-),{\rm in}|
  j_\mu^{em}(0) |0\rangle
  &=&
  +i(p_+-p_-)_\mu F_\pi(s-i\varepsilon),
  \nn\\
  \langle\pi^+({\bf p}_+)\pi^-({\bf p}_-),{\rm out}| j_\mu^{em}(0)
  |0\rangle
  &=&
  -i(p_+-p_-)_\mu F_\pi(s+i\varepsilon),
\end{eqnarray}
with $p_\pm=(E_\pm,{\bf p}_\pm)$ the four-momenta of $\pi^\pm$
and $s=(p_++p_-)^2$ an invariant mass square of the two-pion system. 
The $\pi$-state is normalized as 
\begin{equation}
  \langle\pi^a({\bf p})|\pi^b({\bf q})\rangle
  =
  2E(2\pi)^3 \delta_{ab} \delta({\bf p}-{\bf q}),
  \quad a,b=+,-,0.
\end{equation}
The hadronic electromagnetic current $j_\mu^{em}$ is given in terms of
three-flavor currents as 
$j_\mu^{em} = \frac{2}{3}\bar{u}\gamma_\mu u 
- \frac{1}{3}\bar{d}\gamma_\mu d
-\frac{1}{3}\bar{s}\gamma_\mu s$, 
where $u$, $d$, and $s$ refer to the quark fields. 
One can also write $j_\mu^{em}$ in an isospin basis as 
$j_\mu^{em} = j_\mu^{I=1} +\frac{1}{3}j_\mu^{I=0}
-\frac{1}{3}j_\mu^s$, 
with
\begin{eqnarray}
  j_\mu^{I=1} & = & 
  \frac{1}{2}\left(\bar{u}\gamma_\mu u-\bar{d}\gamma_\mu d\right),
  \nonumber\\
  j_\mu^{I=0} & = &
  \frac{1}{2}\left(\bar{u}\gamma_\mu u+\bar{d}\gamma_\mu d\right),
  \nonumber\\
  j_\mu^s & = &
  \bar{s}\gamma_\mu s.
\end{eqnarray}
In the isospin symmetry limit, the $j_\mu^{I=0}$ and $j_\mu^s$ do not
contribute to $F_\pi(s)$.
Our calculation is performed in the limit of $m_u=m_d$;
thus the vector current is given by $j^{I=1}_\mu$ 
and the $\rho$-$\omega$ mixing effects are neglected. 
To extend the calculation beyond the isospin-symmetric limit, 
the disconnected diagrams need to be calculated, which is a subject of
future studies. 

The pion form factor $F_\pi(s)$ is analytic in the complex $s$-plane,
with a branch cut from $4m_\pi^2$ to $\infty$.
The unitarity of the scattering matrix implies
\begin{equation}
  \label{eq:unitarity}
  \langle f,{\rm out}|j_\mu|0\rangle-
  \langle f,{\rm in}|j_\mu|0\rangle
  =
  -\sum_n
  \left[\langle f,{\rm in}|n,{\rm out}\rangle
    -\delta_{fn}\right]
  \langle n,{\rm out}|j_\mu|0\rangle,
\end{equation}
where $|f\rangle$ stands for the $\pi\pi$ states. 
In the elastic scattering region, due to the energy-momentum
conservation, the sum over $|n\rangle$ is restricted to $\pi\pi$
states as well. 
The coefficient 
$(\langle f,{\rm in}|n,{\rm out}\rangle-\delta_{fn})$ 
is then given by the $\pi\pi$ scattering amplitude.
In the isovector channel, only the $P$-wave amplitude
$t_1(s)=(e^{2i\delta_1(s)}-1)/2i$ contributes to the unitarity
condition, where $\delta_1(s)$ is the $P$-wave $\pi\pi$ scattering
phase. 
One can then simplify (\ref{eq:unitarity}) as
\begin{equation}
  \label{eq:phase}
  {\rm Im} \, F_\pi(s) = t_1^*(s) F_\pi(s+i\varepsilon) =
  \sin\delta_1(s) e^{-i\delta_1(s)} F_\pi(s+i\varepsilon)
\end{equation}
for $s<(4m_\pi)^2$.
It shows that the complex phase of the pion form factor is equivalent
to the $P$-wave $\pi\pi$ scattering phase below the inelastic
threshold. 
This result is known as Watson's final-state theorem.

At low energies the process of $P$-wave $\pi\pi$ scattering is
approximated well by the production and decay of the $\rho$-meson,
which is represented by a simple vector-meson-dominance (VMD) form 
\begin{equation}
  \label{eq:simple_VMD}
  F_\pi^{VMD}(s)=\frac{A}{s-m_\rho^2},\quad A=-m_\rho^2,
\end{equation}
with $m_\rho$ the $\rho$-meson mass. 
The form factor is normalized such that $F_\pi^{VMD}(0)=1$, which is
required by the charge conservation. 
This form is, however, not very satisfactory
since the instability of the $\rho$-meson is not taken into account.
To include the $\pi\pi$ branch cut, Gounaris and Sakurai (GS) 
introduced an analytic form that takes account of the 
$\rho\rightarrow\pi\pi$ transition \cite{Gounaris:1968mw}  
\begin{equation}
  \label{eq:simple_GS}
  F_\pi^{GS}(s) = \frac{A}{s-m_\rho^2-\Pi_\rho(s)},
  \quad A=-m_\rho^2-\Pi_\rho(0),
\end{equation}
where the function $\Pi_\rho(s)$ stands for the $\rho$ meson
self-energy due to the two-pion loop diagram.

Near the resonance energy, the $\rho\rightarrow\pi\pi$
transition amplitude can be parametrized as 
\begin{equation}
  \label{eq:g_rhopipi0}
  \langle\pi^+\pi^-,{\rm out}|\rho,\varepsilon,{\rm in}\rangle = 
  g_{\rho\pi\pi}\,\varepsilon_\mu\cdot(p_+-p_-)^\mu,
\end{equation}
through which the $\rho\pi\pi$ coupling $g_{\rho\pi\pi}$ is defined.
The value of $g_{\rho\pi\pi}$ can be estimated with the experimental
measurement of the $\rho\rightarrow\pi\pi$ decay width
\begin{equation}
  \label{eq:g_rhopipi}
  \Gamma_{\rho\pi\pi} =
  \frac{g_{\rho\pi\pi}^2}{6\pi}\frac{k_\rho^3}{m_\rho^2},
  \quad k_\rho=\sqrt{m_\rho^2/4-m_\pi^2}.
\end{equation}
Using the optical theorem, the imaginary part of $\Pi_\rho(s)$ can be
related to the $\rho\rightarrow\pi\pi$ amplitude, or equivalently
$g_{\rho\pi\pi}$, through  
\begin{equation}
  \label{eq:im_Pi}
  {\rm Im}\,\Pi_\rho(s) = -\frac{g_{\rho\pi\pi}^2}{6\pi}
  \frac{k^3}{\sqrt{s}},
  \quad k=\sqrt{s/4-m_\pi^2}.
\end{equation}
The real part of $\Pi_\rho(s)$ can be related to its imaginary part
using a twice-subtracted dispersion relation. 
Hence, $F_\pi^{GS}(s)$ has only two parameters
$m_\rho$ and $g_{\rho\pi\pi}$.
An explicit expression $F_\pi^{GS}(s)$ is given in
Appendix~\ref{sec:GS_model}.
In particular, the $s$ dependence of the $P$-wave pion-pion scattering phase
induced from the GS model is given in (\ref{eq:phase_shift}). 

\begin{figure}[tb]
  \begin{center}
   \hspace{-15pt}
   \includegraphics[width=280pt,angle=\plotangle]{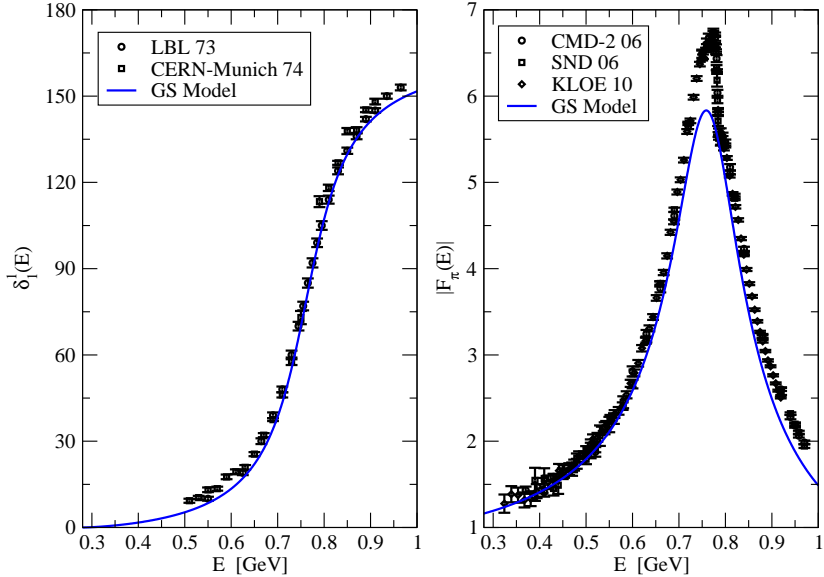}
  \end{center}
  \caption{
    Comparison of the GS model with the experimental measurements of
    $P$-wave pion-pion scattering phase $\delta_1(s)$ and the modulus of
    the pion form factor $|F_\pi|$. 
    We use $E=\sqrt{s}$ as the label of the x-axis. 
    On the left-hand side, circles are from
    \cite{Protopopescu:1973sh}, 
    where the scattering phase is extracted from the reactions
    $\pi^+p\rightarrow\pi^+\pi^-\Delta^{++}$, 
    while the squares from \cite{Estabrooks:1974vu} are based on
    $\pi^-p\rightarrow\pi^-\pi^+n$.
    On the right-hand side, circles, squares and diamonds stand for
    the data of $|F_\pi(s)|$, 
    compiled using 
    the CMD-2 06~\cite{Akhmetshin:2006wh,Akhmetshin:2006bx}, 
    SND 06~\cite{Achasov:2006vp}, 
    and KLEO 10 $e^+e^-$ data~\cite{Ambrosino:2010bv}, respectively. 
    The blue curve shows the GS model (\ref{eq:phase_shift}) and
    (\ref{eq:GS_FF}), 
    where the Particle Data Group (PDG)~\cite{Beringer:1900zz} values
    $m_\pi$ = 0.1395702(4)~GeV and $m_\rho$ = 0.7753(3)~GeV are inputs 
    and $g_{\rho\pi\pi}$ = 5.95(2) is estimated with the PDG value of
    $\Gamma_\rho$ = 0.1478(9)~GeV.
  }
  \label{fig:GS_compare}
\end{figure}

As shown in Fig.~\ref{fig:GS_compare}, the GS form gives a
reasonably good approximation of the experimental measurements of the
scattering phase, but $|F_\pi^{GS}(s)|$ is about 10\% lower near the 
resonance peak $\sqrt{s}=m_\rho$.
The deviation may arise from the $\rho-\omega$ mixing due to the
isospin breaking effect.
In \cite{Jegerlehner:2011ti} the $\omega$ contribution is subtracted
from the CMD-2 data and the peak value of the form factor is only
$\sim$~3\% smaller than the original one, which suggests that
the $\rho-\omega$ mixing effect is not the only source of the deviation
between the GS model and experimental data. 
This is further confirmed by our lattice calculation, where 
the up and down quark masses are set identical while the peak value of
the GS form factor is 27\% and 20\% smaller than the lattice results
at $m_\pi=380$ MeV and 290 MeV, respectively, as shown later in
Fig.~\ref{fig:form_factor}. 

One way to make the GS form closer to the experimental data is to
include the contributions from higher resonances such as $\rho(1450)$
and $\rho(1700)$ \cite{Schael:2005am,Davier:2005xq}. 
After doing this, the extended GS form does agree with the
experimental measurements but there are still some doubts on whether
the higher resonances really affect the form factor at the
$\rho$-resonance peak in the suggested way \cite{Jegerlehner:2011ti}.

Another way to modify the GS form is to focus only on the resonance
region $s\approx m_\rho^2$ and assume the $\rho$-meson dominance. 
The matrix elements in (\ref{eq:matrix_element}) are then factorized
into two parts: 
$\langle\pi^+\pi^-,{\rm out}|\rho,\varepsilon,{\rm in}\rangle$ and
$\langle\rho,\varepsilon,{\rm in}|j_\mu|0\rangle =
g_{\rho,em}m_\rho^2\varepsilon_\mu$, 
where the former one is related to $g_{\rho\pi\pi}$ by
(\ref{eq:g_rhopipi0}) and the latter yields the $\rho$-meson decay
constant $g_{\rho,em}$. 
Consequently, the form factor is constructed as \cite{Melikhov:2002nf,Bruch:2004py}
\begin{equation}
  \label{eq:GS_VMD}
  F_{\pi}^{GS+VMD}(s) =
  \frac{A}{s-m_\rho^2-\Pi_\rho(s)},
  \quad 
  A=-g_{\rho\pi\pi} g_{\rho,em} m_\rho^2,
\end{equation}
where the numerator is given by $-g_{\rho\pi\pi} g_{\rho,em} m_\rho^2$ and 
the denominator still uses the dressed $\rho$ propagator.
Using $g_{\rho\pi\pi}=5.95(2)$ and $g_{\rho,em}=0.2017(9)$ extracted
from the $\rho\rightarrow e^+e^-$ decay width as inputs, this formula
gives a good description of the experimental data near the resonance
peak but violates the charge conservation condition at $s=0$.

Comparing (\ref{eq:simple_GS}) to (\ref{eq:GS_VMD}), it is natural to
introduce an $s$-dependent $A(s)$ and write the form factor as
\begin{equation}
  \label{eq:ff}
  F_\pi(s)   = \frac{A(s)}{s-m_\rho^2-\Pi_\rho(s)} =  F_\pi^{GS}(s)
  \sum_{n=0}^N c_n(s-m_\rho^2)^n.
\end{equation}
Here we use a Taylor expansion at $s=m_\rho^2$ to describe the
behavior of the form factor near the resonance region.
The polynomial terms are introduced to account for the deviation
between the $F_\pi^{GS}(s)$ given by (\ref{eq:simple_GS}) and
the $I=1$ part of the experimental data, 
which may arise from the interference between $\rho$ and higher
resonances such as $\rho(1450)$ and $\rho(1700)$.
The coefficients $c_n$ should respect the charge conservation
condition, {\it i.e.} $\sum_{n=0}^{N} c_n(-m_\rho^2)^n=1$.

In our work, since we calculate the scattering phase and the modulus
of the form factor at several discrete energies, we adopt the form
(\ref{eq:ff}) to describe their $s$ dependence. 
This induces a model dependence in our final results for the parameters 
$m_\rho$, $g_{\rho\pi\pi}$ and the charge radius $\langle r_\pi^2\rangle$.
But the model dependence will become milder if one collects more data
points at various energies.
As the data points become dense, lattice QCD will eventually provide a
complete description of the low-energy timelike pion form factor from
the first principles.

\section{Finite-size method}
\label{sect:FSM}
According to the general idea of \cite{Luscher:1990ux} for the study
of the two-body scattering problem on the lattice, we consider 
the two-pion system in a box of finite size $L$.

Given an $I=1$ vector-current operator 
$j_{\bf b}=\bar{\psi}({\bf b}\cdot\gamma)\frac{\tau^3}{2}\psi$ 
one can construct a correlation function in a finite volume $V=L^3$ as
\begin{equation}
  C_V(t) =
  \int_Vd^3{\bf x} \;e^{-i{\bf P}\cdot{\bf x}}
  \langle0|j_{\bf b}({\bf x},t)j_{\bf b}^\dagger({\bf 0},0)|0\rangle,
\end{equation}
where a unit vector ${\bf b}$ indicates the polarization direction of
the vector current and ${\bf P}$ is the total three-momentum. 
When ${\bf P}\neq{\bf 0}$, {\bf b} can be set either parallel or
perpendicular to ${\bf P}$ to make the operator $j_{\bf b}$ belong to a
certain irreducible representation of the rotational group. 
Since $j_{\bf b}$ has the same quantum number as a two-pion system in
the $I=1$ channel, two-pion states appear in the correlator as
intermediate states,
\begin{equation}
  \label{eq:corr_finite}
  C_V(t)
  \rightarrow 
  \sum_n |\langle0|j_{\bf b}|\pi\pi,n\rangle_V|^2 e^{-E_nt}.
\end{equation}
Here the arrow denotes the asymptotic contributions in the large time
separations, where the $\pi\pi$ states of various relative momenta
dominate as the lowest energy states. 

By studying the time dependence of the correlator, one obtains two
observables from (\ref{eq:corr_finite}): 
$E_n$ and $|\langle0|j_{\bf b}|\pi\pi,n\rangle_V|^2$.
The discrete energy $E_n$ contains the information of pion-pion
scattering and can be related to the infinite-volume $P$-wave
scattering phase $\delta_1$ by the L\"uscher formula \cite{Luscher:1990ux}
and its extension to the moving frames where the total momenta 
${\bf P}$ is nonzero
\cite{Rummukainen:1995vs,Christ:2005gi,Kim:2005gf} 
\begin{equation}
  \label{eq:Luscher_formula}
  n\pi-\delta_1(k) =
  \phi^{{\bf P},\Gamma}(q=kL/2\pi),
  \quad
  \sqrt{s}=\sqrt{E_n^2-{\bf P}^2}=2\sqrt{m_\pi^2+k^2}.
\end{equation}
Here, $\phi^{{\bf P},\Gamma}(q)$ is a known function, irrelevant to
the details of the interaction.
It only depends on the moving frame ${\bf P}$ and the irreducible
representation $\Gamma$ that the operator $j_{\bf b}$ belongs to.
The ``momentum'' $k$ entering in $\phi^{{\bf P},\Gamma}(q)$ through
$q$ is indirectly determined by the energy $E_n$ as shown by the
second equation of (\ref{eq:Luscher_formula}).
The formula (\ref{eq:Luscher_formula}) is widely used in various
lattice calculations of the $P$-wave pion-pion scattering phase and
the studies of the $\rho$-resonance properties. 
The formulas used in this calculation are listed in Appendix~\ref{sect:Luscher_formula}.

Since $E_n$ can be used to determine the scattering phase $\delta_1$, 
which is the complex phase of $F_\pi(s)$, 
a natural question arises whether one can relate 
$|\langle0|j_{\bf b}|\pi\pi,n\rangle_V|^2$ to $|F_\pi(s)|^2$.
Meyer gave an answer to this question in \cite{Meyer:2011um}, where he
introduced an external vector particle $W$ which couples to the quarks
via an infinitesimal interaction 
$H_{\rm int}(x)=ej_\mu(x)W^\mu(x)$. 
Then, the matrix element $\langle\pi\pi,{\rm out}|j_\mu(0)|0\rangle$
is related to the amplitude 
$\langle\pi\pi,{\rm out}|H_{\rm int}(0)|W\rangle$, 
which is analogous to the $K\rightarrow\pi\pi$ transition amplitude
$\langle\pi\pi,{\rm out}|{\mathcal L}_W(0)|K\rangle$. 
The techniques used in deriving the Lellouch-L\"uscher formula for
$K\rightarrow\pi\pi$ \cite{Lellouch:2000pv} can thus be transplanted
to the case of $W\rightarrow\pi\pi$. 
The main difference is that $K\rightarrow\pi\pi$ contains an $S$-wave
$\pi\pi$ scattering in the $I=0$ or 2 channel 
while $W\rightarrow\pi\pi$ has a $P$-wave scattering in the $I=1$
channel. 
We generalize the formula of \cite{Meyer:2011um} to the case of
general moving frames.
The relation between the finite-volume matrix element 
$|\langle0|j_{\bf b}|\pi\pi,n\rangle_V|^2$ 
and the square of the modulus of the form factor in the infinite
volume is written as
\begin{equation}
  \label{eq:LL_formula}
  |F_\pi(s)|^2 = \frac{\gamma}{g(\gamma)^2}
  \left( k\frac{\partial\delta_1(k)}{\partial k} +
    q\frac{\partial \phi^{{\bf P},\Gamma}(q)}{\partial q} \right)
  \frac{3\pi s}{2k^5}
  |\langle0|j_{\bf b}(0)|\pi\pi,n\rangle_V|^2,
\end{equation}
where $s$ takes the discrete values $s=E_n^{*2}$ with $E_n^*$ the
center-of-mass energy of the state corresponding to $E_n$.
$\gamma$ is a Lorentz boost factor $\gamma=E_n/E_n^*$ and the function
$g(\gamma)$ takes the value of 
$g(\gamma)=\gamma$ for ${\bf b}\parallel{\bf P}$ and 
$g(\gamma)=1$ for ${\bf b}\perp{\bf P}$.
In the case of vanishing ${\bf P}$, (\ref{eq:LL_formula}) reduces to
the formula in \cite{Meyer:2011um}.

In the $K\rightarrow\pi\pi$ decays, the power-law finite-volume
corrections are accounted for by the $\pi\pi$-states rather than the
single $K$-states.
It is therefore simpler to retain the essential physical aspects of 
$\pi\pi$ and eliminate the kaon \cite{Lin:2001ek}.
Following this idea, we make another demonstration of
(\ref{eq:LL_formula}) without introducing the fictitious state $W$. 
Some details are given in Appendix~\ref{sect:LL_Pwave}.

\section{Lattice setup}
\label{sect:corr_func}
In this work we use the $2+1$-flavor overlap fermion ensembles
generated by the JLQCD Collaboration \cite{Aoki:2012pma,Fukaya:2006vs}.
Using the overlap fermions ensures exact chiral symmetry in the chiral
limit at finite lattice spacings.
The calculation is performed at bare quark masses $am$ = 0.025 and
0.015, that correspond to the pion masses 
$m_\pi$ = 380 and 290~MeV, respectively.
Physical kinematics that the $\rho$ meson decays to two pions is
realized in both cases.
The Iwasaki gauge action is employed together with the unphysically
heavy Wilson fermions that prevent the topological charge from
changing its value during the hybrid Monte Carlo simulation \cite{Fukaya:2006vs}.
The $\beta$ value is 4.30, that corresponds to the lattice spacing $a$
= 0.112(1)~fm for both pion masses. To take full control of
systematic effects, having multiple lattice spacings and performing a
continuum extrapolation are important. This would require further
simulation efforts and shall be done in the future.
The lattice size is $(L/a)^3\times(T/a)=24^3\times48$, and 
the lattice extent $L$ in the physical unit is 2.6~fm, 
which roughly satisfies $m_\pi L\gtrsim 4$.
The effect of fixing topological charge would not be significant on
such large volume lattice \cite{Aoki:2007ka}.

We construct a vector-current operator using two-flavor quark fields 
$\bar{\psi}$ and $\psi$ and consider its Fourier transform 
\begin{equation}
  \label{eq:operator_quark}
  j_{\bf b}^{\bar{\psi}\psi}({\bf P},t) = 
  \frac{Z_V}{L^{3/2}}
  \sum_{\bf x}e^{-i{\bf P}\cdot{\bf x}}
  \left( \bar{\psi}({\bf b}\cdot\gamma)\frac{\tau^3}{2}\psi \right)
  ({\bf x},t),
\end{equation}
where ${\bf b}$ is a unit vector and ${\bf b}\cdot\gamma$ is defined as
\begin{equation}
  {\bf b}\cdot\gamma = \sum_{i=1}^3 b_i\Gamma_i^{\rm rot},
  \quad 
  \Gamma_i^{\rm rot} = \gamma_i\left(1-\frac{aD_{ov}(0)}{2m_0}\right).
\end{equation}
Here, we use the rotated gamma matrices $\Gamma_i^{\rm rot}$ to remove
the $O(a)$ lattice artifacts from the interpolating operator.
$D_{ov}(m_q)$ is the overlap-Dirac operator for the quark mass $m_q$,
and $m_0=1.6$ is the (negative) mass parameter to define the kernel of
the overlap-Dirac operator. 
In the continuum limit $a=0$, $\Gamma_i^{\rm rot}$ reduces to the
conventional gamma matrix $\gamma_i$. 
$Z_V$ is the renormalization factor for the vector currents. 
Its value $Z_V=1.39360(48)$ is obtained nonperturbatively \cite{Noaki:2009xi}.

Besides the construction using the quark fields, one can also define
the vector-current operator using $\pi^+\pi^-$ meson pairs
\begin{equation}
  \label{eq:operator_pi0}
  j_{\bf b}^{(\pi\pi,n)}({\bf P},t) =
  \pi^+({\bf p}_1,t)\pi^-({\bf p}_2,t) -
  \pi^+({\bf p}_2,t)\pi^-({\bf p}_1,t),
\end{equation}
where the pion interpolating operator $\pi^\pm({\bf p},t)$ is defined as
\begin{equation}
  \pi^\pm({\bf p},t) =
  \frac{1}{L^{3/2}} \sum_{\bf x} e^{-i{\bf p}\cdot{\bf x}}
  \left( \bar{\psi}\Gamma_5^{\rm rot}\frac{\tau^\pm}{2}\psi \right)
  ({\bf x},t).
\end{equation}
The momenta ${\bf p}_{1,2}$ satisfy 
$\frac{L}{2\pi}{\bf p}_{1,2}\in{\mathbb Z}^3$.
The total three-momentum of the two-pion system is given by 
${\bf P}={\bf p}_1+{\bf p}_2$ 
and the polarization direction is defined as 
${\bf b}=\frac{{\bf p}_1-{\bf p}_2}{|{\bf p}_1-{\bf p}_2|}$.
The index $n$ specifies the energy levels 
corresponding to
$E_n=\sqrt{m_\pi^2+{\bf p}_1^2}+\sqrt{m_\pi^2+{\bf p}_2^2}$.

We can modify the two-pion interpolating operator
(\ref{eq:operator_pi0}) by 
separating the two pion-operators at different time slices
\begin{eqnarray}
  \label{eq:operator_pi}
  j_{\bf b}^{(\pi\pi,n)}({\bf P},t) & = &
  \frac{1}{2}\left[
    \pi^+({\bf p}_1,t_1)\pi^-({\bf p}_2,t_2) +
    \pi^+({\bf p}_1,t_2)\pi^-({\bf p}_2,t_1)
  \right]\nn\\
  & - &
  \frac{1}{2}\left[
    \pi^+({\bf p}_2,t_1)\pi^-({\bf p}_1,t_2) +
    \pi^+({\bf p}_2,t_2)\pi^-({\bf p}_1,t_1)
  \right],
  \quad t_{1,2}=t\pm \delta t.
\end{eqnarray}
By swapping ${\bf p}_{1,2}\rightarrow{\bf p}_{2,1}$ or
$\pi^\pm\rightarrow\pi^{\mp}$ 
we have $j_{\bf b}^{(\pi\pi,n)}\rightarrow -j_{\bf b}^{(\pi\pi,n)}$, 
which verifies that the operator defined in (\ref{eq:operator_pi}) is parity-odd and isospin-odd. 
The reasons to use (\ref{eq:operator_pi}) in our calculation are
twofold:
First, we use the all-to-all propagator \cite{Foley:2005ac}
in our calculation. 
When the two pions are put on the same time slice, a different
stochastic source for each pion is required to avoid unphysical
contributions, but in our implementation \cite{Aoki:2009qn},
only one stochastic source is used for each time slice.
Therefore we separate the two pions at different time slices to avoid
the unwanted contributions.
Second, by separating with a distance of $2\delta t$, the correlation
between the two pion-operators is reduced. 
As a consequence, the precision of the correlator can be improved.
For example, in the case of ${\bf P}={\bf 0}$, the error of the
effective energy is reduced by a factor of 3 by introducing a
separation of $\delta t/a=1$. 
We examine also the case of $\delta t/a=2$ and $3$, but the change is
not very significant.
A drawback of using a large $\delta t$ is that it enhances the
excited-state effects because the 
minimum time separation between pion fields in $j_{\bf b}^{(\pi\pi,n)}({\bf P},t)$ 
and $j_{\bf b}^{(\pi\pi,n)}({\bf P},0)$ is
$t-2\delta t$ rather than $t$.
In this calculation we simply use $\delta t/a=1$.
As indicated in \cite{Liu:2011jp}, separating the two pion-operators
can also be useful in the calculation of the $I=0$ pion-pion
scattering, where it reduces the noise dramatically from the
disconnected diagram.

With the vector-current operator $j_{\bf b}^{\bar{\psi}\psi}$ or 
$j_{\bf b}^{(\pi\pi,n)}$, one can construct operators in the
irreducible representations of the cubic group (and reflections)
using the standard procedure of the character projection
\begin{equation}
  \label{eq:operator_symmetry}
  j^{q}(\Gamma,{\bf P},t) =
  \frac{d_\Gamma}{N_G}
  \sum_{\hat{R}\in G}
  \chi_\Gamma^*(\hat{R})j_{\hat{R}\bf b}^{q}({\bf P},t),
\end{equation}
where $q=\bar{\psi}\psi$ or $(\pi\pi,n)$,
and $N_G=\sum_{\hat{R}\in G}1$.
The notations follow those of \cite{Feng:2010es,Feng:2011ah}.
Here the symmetry group $G$ is introduced as the set of all lattice
rotations and reflections $\hat{R}$. 
In the case of ${\bf P}={\bf 0}$, $G$ reduces to the full cubic group 
$O_h$. 
For ${\bf P}\neq{\bf 0}$, on the other hand, $G$ spans a subspace of
$O_h$, under which the momentum ${\bf P}$ is invariant or changes only
by a minus sign
\begin{equation}
  \label{eq:group}
  G = \left\{
    \hat{R}\in O_h\bigg | 
    \hat{R}{\bf P}={\bf P}\,\,
    \text{or}\,\,
    \hat{R}{\bf P}=-{\bf P}
  \right\}.
\end{equation}
$\Gamma$ is the irreducible representation of the group $G$, while
$d_\Gamma$ and $\chi_\Gamma(\hat{R})$ are the dimension and character
of $\Gamma$, respectively.
The character projection makes the operator $j^{q}(\Gamma,{\bf P},t)$ belong to 
a given representation $\Gamma$.

\begin{table}
  \centering
  \begin{tabular}{c||ccccc}
    \hline
    No. & ${\bf P}$ & $G$ & $\Gamma$ & 
    $j^{(\pi\pi,n)}_{\bf b}$: [${\bf p}_1$, ${\bf p}_2$] & 
    $j^{\bar{\psi}\psi}_{\bf b}$: ${\bf b}$ \\
    \hline
    \multirow{3}{*}{\textcircled{1}} & \multirow{3}{*}{$(0,0,0)$} &
    \multirow{3}{*}{$O_h$} & \multirow{3}{*}{$T_1^-$}  &  
    [$(1,0,0)$, $(-1,0,0)$] &  $(1,0,0)$ \\
    &		    & & &  [$(0,1,0)$, $(0,-1,0)$] &  $(0,1,0)$ \\
    &		    & & &  [$(0,0,1)$, $(0,0,-1)$] &  $(0,0,1)$ \\
    \textcircled{2} & $(0,0,1)$  & $D_{4h}$ & $A_2^-$ 
    & [$(0,0,1)$, $(0,0,0)$] &  $(0,0,1)$ \\
    \textcircled{3} & $(1,1,0)$  & $D_{2h}$ & $B_1^-$ 
    & [$(1,1,0)$, $(0,0,0)$] &  $\frac{1}{\sqrt{2}}(1,1,0)$ \\
    \textcircled{4} & $(1,1,1)$  & $D_{3d}$ & $A_2^-$ 
    & [$(1,1,1)$, $(0,0,0)$] &  $\frac{1}{\sqrt{3}}(1,1,1)$ \\
    \textcircled{5} & $(1,1,0)$  & $D_{2h}$ & $B_2^-$ 
    & [$(1,0,0)$, $(0,1,0)$] &  $\frac{1}{\sqrt{2}}(1,-1,0)$ \\
    \hline
  \end{tabular}
  \caption{
    \textcircled{1}, ..., \textcircled{5} identify the operators used
    in this calculation. 
    ${\bf P}$ denotes the total three-momentum in units of $2\pi/L$. 
    $G$ is the cubic rotational group defined in (\ref{eq:group}). 
    Since the reflection operator is involved, $G$ is a parity doubled
    little group associated with momentum ${\bf P}$. 
    $\Gamma$ stands for the irreducible representation of group $G$. 
    $T_1^-$ is a three-dimensional representation while others are
    one dimensional. 
    For a given $\Gamma$, one can construct the operators using
    (\ref{eq:operator_symmetry}). 
    In our calculation, these interpolating operators can be
    simplified as 
    $j^{(\pi\pi,n)}_{\bf b}$ and $j^{\bar{\psi}\psi}_{\bf b}$. 
    The $j^{(\pi\pi,n)}_{\bf b}$ are specified using the momenta 
    ${\bf p}_1$ and ${\bf p}_2$ in units of $2\pi/L$. 
    The $j^{\bar{\psi}\psi}_{\bf b}$ can be determined by the
    polarization ${\bf b}$. 
    Note that, although the operators \textcircled{1} and
    \textcircled{2} contain the $j^{\bar{\psi}\psi}_{\bf b}$ with the
    same polarization ${\bf b}=(0,0,1)$, 
    the different total momentum ${\bf P}$ makes them belong to the
    different representations of different groups.
  }
  \label{tab:operator}
\end{table}

In a general moving frame with nonzero ${\bf P}$, the operator
$j_{\bf b}^{\bar{\psi}\psi}$ with ${\bf b}\parallel{\bf P}$ forms a
basis of a one-dimensional representation of $G$.
For the operators belonging to the other representations,
we take ${\bf b}$ and ${\bf P}$ such that ${\bf b}\perp{\bf P}$.
In general, $j^{q}(\Gamma,{\bf P},t)$ defined in
(\ref{eq:operator_symmetry}) is a linear combination of a few
$j_{\bf b}^q$ with different polarization ${\bf b}$, but
with our choice these interpolating operators can be simply given by a
single $j_{\bf b}^q$.
We list the operators used in our calculation in Table~\ref{tab:operator}.

\begin{figure}[tb]
  \begin{center}
   \hspace{-15pt}
   \includegraphics[width=380pt,angle=0]{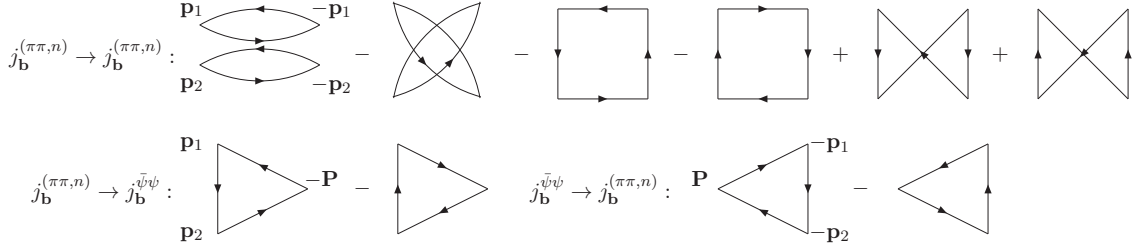}
  \end{center}
  \caption{Quark contractions for three- and four-point correlation functions. 
  The momenta $\pm {\bf p}_{1,2}$ are used to indicate the
  single pion field. $\pm{\bf P}$ are used to specify the $j_{\bf b}^{\bar{\psi}\psi}$ field.}
  \label{fig:quark_flow}
\end{figure}

Using the operators \textcircled{1}, ..., \textcircled{5} in
Table~\ref{tab:operator}, for each set of 
$\{j_{\bf b}^{\bar{\psi}\psi},j_{\bf b}^{(\pi\pi,n)}\}$, 
we can construct a $2\times2$ correlation matrix with its matrix
elements defined through
\begin{equation}
  \label{eq:correlation_matrix}
  C_{q,q'}(t) = \frac{1}{T}\sum_{t_0=0}^{T-1}
  \left\langle 
    j_{\bf b}^{q}({\bf P},t+t_0)\,
    j_{\bf b}^{q'}({\bf P},t_0)^\dagger
  \right\rangle,
  \quad 
  q,q'=\bar{\psi}\psi\,\text{or}\,(\pi\pi,n).
\end{equation}
The quark contractions for three- and four-point correlation functions are shown in Fig.~\ref{fig:quark_flow}
Then the variational method \cite{Luscher:1990ck} allows us to isolate
the ground state and first excited state from the correlation matrix. 
From each of the five operator sets, we can calculate two energy
eigenvalues, so that we obtain the scattering phase and the pion form
factors at ten discrete energies.
As shown in (\ref{eq:correlation_matrix}), we perform a time
translation average to reduce the statistical noise of the correlators. 
This requires the quark propagator inversions at each time slice. 
For ${\bf P}={\bf 0}$, we average the correlators using the three
operator sets in \textcircled{1}, since $T_1^-$ is a three-dimensional
representation. 
For ${\bf P}\neq{\bf 0}$ we average the correlators carrying total
momentum ${\bf P}$ with those carrying momenta 
$\hat{R}{\bf P}$ ($\hat{R}\in O_h$), 
since these correlators are equivalent under the symmetry. 
This requires various momentum insertions in the propagator
inversions. 
All these requirements are fulfilled by using the all-to-all
propagators generated by the JLQCD Collaboration.

Here we briefly describe the construction of the all-to-all propagator
\cite{Foley:2005ac,Bali:2005fu} by the JLQCD Collaboration \cite{Aoki:2009qn}.
The quark propagator $D^{-1}(x,y)$ can be explicitly composed using the eigenmodes
of the Hermitian Dirac operator
\ba
&& H(x,y)=\gamma_5 D(x,y),\quad D^{-1}(x,y)=H^{-1}(x,y)\gamma_5,
\nn\\
&&H(x,y)u_n(y)=\lambda_n u_n(x)\quad\Rightarrow\quad H^{-1}(x,y)=\sum_n \frac{1}{\lambda_n}u_n(x)u_n^\dagger(y),
\ea 
where $H(x,y)$ is a Hermitian matrix with its color and spinor indices omitted for simplicity.  
$\lambda_n$ is the $n_{\rm th}$ eigenvalue and $u_n(x)$ the associated eigenvector.
However, it is not realistic to calculate all the eigenmodes. 
So we decompose the propagator into low- and high-mode contributions using a projection operator
$P_{\rm low}(x,y)=\sum_{n=1}^{N_\lambda}u_n(x)u_n^\dagger(y)$
\ba
H^{-1}(x,y)&=&H^{-1}_{\rm low}(x,y)+H^{-1}_{\rm high}(x,y),
\nn\\
H^{-1}_{\rm low}(x,y)&=&H^{-1}(x,z)P_{\rm low}(z,y)=\sum_{n=1}^{N_\lambda} \frac{1}{\lambda_n}u_n(x)u_n^\dagger(y),
\nn\\
H^{-1}_{\rm high}(x,y)&=&H^{-1}(x,z)(\delta_{z,y}-P_{\rm low}(z,y)).
\ea
We use only the low-lying eigenmodes and supplement them with the
remaining high-mode contributions calculated with a stochastic
method
\ba
H(x,y)\phi_{r,d}(y)=(\delta_{x,z}-P_{\rm low}(x,z))\eta_{r,d}(z)\quad
\Rightarrow\quad H^{-1}_{\rm high}(x,y)=\frac{1}{N_r}\sum_{r=1}^{N_r}\sum_{d=1}^{N_d}\phi_{r,d}(x)\eta_{r,d}^\dagger(y),
\ea
where $r=1,\cdots,N_r$ indicates the complex $Z_2$ stochastic sources and $d=1,\cdots N_d$ specifies the dilutions in spin, color and space-time positions.
Combining the low modes and high modes together yields the so-called
all-to-all propagator. 
In our analysis we use 50 configurations for each ensemble. For each configuration, we use $N_\lambda=240$, $N_r=1$ and
$N_d=3\times4\times T/2=288$. 
For more details of the all-to-all propagator technique, we refer
readers to \cite{Foley:2005ac,Bali:2005fu,Aoki:2009qn}.

\section{Analysis}
\label{sect:analysis}
\subsection{Removal of the around-the-world effects}
Before applying the variational technique for the sets of correlators,
we first remove the so-called around-the-world effect, which arises
due to the finite time extent $T$ in the lattice calculation. 
This effect modifies the time dependence of single pion correlator
$\langle\pi(t)\pi(0)\rangle$ as $e^{-E_\pi t}+e^{-E_\pi(T-t)}$, with
the around-the-world contribution $e^{-E_\pi(T-t)}$. 
In the calculation of the pion-pion scattering, it can cause a
discernible effect especially near $t\sim T/2$ \cite{Feng:2009ij,Blum:2011pu,Dudek:2012gj}.

To find out how the around-the-world effects deform the correlator, we
insert a complete set of eigenstates into the correlators in
(\ref{eq:correlation_matrix}) as
\begin{eqnarray}
  C_{q,q'}(t) & = &
  \sum_{m,m'}
  \langle m|j_{\bf b}^q|m'\rangle
  \langle m'|j_{\bf b}^{q'\dagger}|m\rangle 
  e^{-E_m(T-t)}e^{-E_{m'}t}
  \nonumber\\
  & = &
  \sum_n 
  \langle 0|j_{\bf b}^q|\pi\pi,n\rangle
  \langle \pi\pi,n|j_{\bf b}^{q'\dagger}|0\rangle  
  \left(e^{-E_{\pi\pi,n}(T-t)}+e^{-E_{\pi\pi,n}t}\right)
  \nonumber\\
  & & +
  \sum_{{\bf p}_1,{\bf p}_2}
  \langle \pi|j_{\bf b}^q|\pi\rangle
  \langle \pi|j_{\bf b}^{q'\dagger}|\pi\rangle 
  \left(
    e^{-E_\pi({\bf p}_2)(T-t)}e^{-E_\pi({\bf p}_1)t} +
    e^{-E_\pi({\bf p}_1)(T-t)}e^{-E_\pi({\bf p}_2)t}
  \right)+\cdots.
  \nonumber\\
\end{eqnarray}
In the last equation, the first term represents the physical
contribution from the lowest energy states 
$|m(m')\rangle=|0\rangle$ and $|m'(m)\rangle=|\pi\pi,n\rangle$.
The second term is the around-the-world contribution, which arises by
setting 
$|m(m')\rangle=|\pi({\bf p}_1)\rangle$ and 
$|m'(m)\rangle=|\pi({\bf p}_2)\rangle$.
Note that the interpolating operator $j_{\bf b}^{q,q'}$ carries a
three-momentum ${\bf P}$. 
The momenta ${\bf p}_1$, ${\bf p}_2$ and ${\bf P}$ satisfy the momentum
conservation. 
The largest contamination thus comes from the terms with 
${\bf p}_1={\bf 0}$ and ${\bf p}_2={\bf P}$ or 
${\bf p}_1=-{\bf P}$ and ${\bf p}_2={\bf 0}$.

To reduce the bulk of these around-the-world effects, we construct a
modified correlator through
\begin{equation}
  \label{eq:modify_corr}
  \bar{C}_{q,q'}(t) =
  C_{q,q'}(t) - C_{q,q'}(t+\Delta t)
  \frac{\cosh\left[\Delta E(T/2-t)\right]}{
    \cosh\left[\Delta E(T/2-(t+\Delta t))\right]},
\end{equation}
where $\Delta E=E_\pi({\bf P})-E_\pi({\bf 0})$.
With too small $\Delta t$ a cancellation between $C_{q,q'}(t)$ and
$C_{q,q'}(t+\Delta t)$ makes the modified correlator noisy, while
too large $\Delta t$ yields larger intrinsic noise due to large time
separation $t+\Delta t$.
As a compromise, we take $\Delta t/a=6$.

\subsection{Extracting the eigenstates}
After removing the around-the-world effects, we apply the variational 
method \cite{Luscher:1990ck} to extract the energy $E_n$ and the
matrix element 
$|\langle0|j_{\bf b}^{q}|\pi\pi,n\rangle_V|^2$ 
from the correlation matrix. 
The procedure is as follows. We first build the correlation matrix using the modified correlator in (\ref{eq:modify_corr}).
By constructing a ratio of the correlation matrix
\begin{equation}
  \label{eq:R_ratio}
  R(t,t_R) = 
  \bar{C}^{-\frac{1}{2}}(t_R)\bar{C}(t)\bar{C}^{-\frac{1}{2}}(t_R),
\end{equation}
and solving the eigensystem of
\begin{equation}
  R(t,t_R)B_n=D_n(t,t_R)B_n,\quad n=0,1
\end{equation}
one can determine the eigenvalues $D_n(t,t_R)$ and the normalized
eigenvectors $B_n$ for $t>t_R$. 
Since $R(t,t_R)$ is a Hermitian matrix, the eigenvectors $B_n$ form an
orthogonal system, {\it i.e.} $B^\dagger B=1$.
Then, $D_n(t,t_R)$ is related to the energy eigenvalues of the $\pi\pi$ scattering states through
\begin{equation}
  D_n(t,t_R)=D_n(t)/D_n(t_R),
\end{equation}
with the function $D_n(t)$ defined as
\begin{equation}
  \label{eq:Dn_def}
  D_n(t) = \left(e^{-E_nt}+e^{-E_n(T-t)}\right)
  \left(1-
    \frac{
      \cosh\left[E_n(T/2-(t+\Delta t))\right] 
      \cosh\left[\Delta E(T/2-t)\right]
    }{
      \cosh\left[E_n(T/2-t)\right]
      \cosh\left[\Delta E(T/2-(t+\Delta t))\right]
    }
  \right).
\end{equation}
Since $\Delta E$ and $\Delta t$ are known, $D_n(t)$ is a function of
only $E_n$ and $t$. 
Using the lattice data of $D_n(t,t_R)$ as inputs, one can determine
$E_n$. 

Note that the eigenvectors of $R(t,t_R)$ can also be given by $\bar{C}^{\frac{1}{2}}(t_R)A^{-1}$, with $A_{n,q}$ defined as
$A_{n,q}=\langle\pi\pi,n|j_{\bf b}^{q\dagger}|0\rangle_V$. 
A relation between $B$ and $A$ is then established through
\begin{equation}
  \label{eq:relation_B_A}
  B_{q,n}  = X_n 
  \left[\bar{C}^{\frac{1}{2}}(t_R)A^{-1}\right]_{q,n}
  \quad
  \Rightarrow
  \quad 
  \left[\bar{C}^{-\frac{1}{2}}(t_R)B\right]_{q,n}
  = X_n\left[A^{-1}\right]_{q,n}
\end{equation}
with a coefficient $X_n$ to be determined. 
$B^\dagger B=1$ leads to $|X_n|^2=D_n^{-1}(t_R)$.
Making use of the relation (\ref{eq:relation_B_A}), we obtain
\begin{eqnarray}
  \label{eq:amplitude}
  &&
  \left[
    B^\dagger \bar{C}^{-\frac{1}{2}}(t_R) \bar{C}(t)
  \right]_{n,q}
  = X_n^* D_n(t)A_{n,q}
  \nonumber\\
  &&
  \hspace{3cm}\Rightarrow\quad
  D_n(t_R)|A_{n,q}|^2 = 
  \left| 
    \left[
      B^\dagger \bar{C}^{-\frac{1}{2}}(t_R) \bar{C}(t)
    \right]_{n,q}
  \right|^2 
  D_n^{-2}(t,t_R).
\end{eqnarray}
Since $D_n(t,t_R)$ and $B$ are known, (\ref{eq:amplitude}) can be used 
to extract $D_n(t_R)|A_{n,q}|^2$. 
By putting the evaluated value of $E_n$ into (\ref{eq:Dn_def}), one
can remove $D_n(t_R)$ and determine $|A_{n,q}|^2$.

\begin{figure}[p]
  \centering
  \includegraphics[width=300pt,angle=270]
  {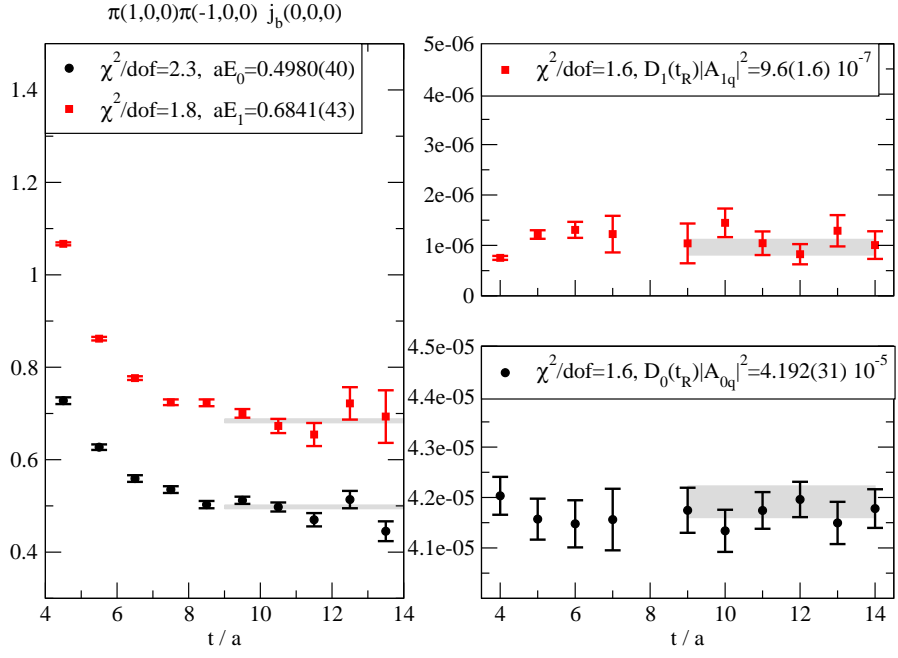}
  \caption{
    Effective energies and amplitudes for the operator set \textcircled{1}
    and $m_\pi$ = 380~MeV. 
  }
  \label{fig:mode1_mpi380}
\end{figure}

\begin{figure}[p]
  \centering
  \includegraphics[width=300pt,angle=270]
  {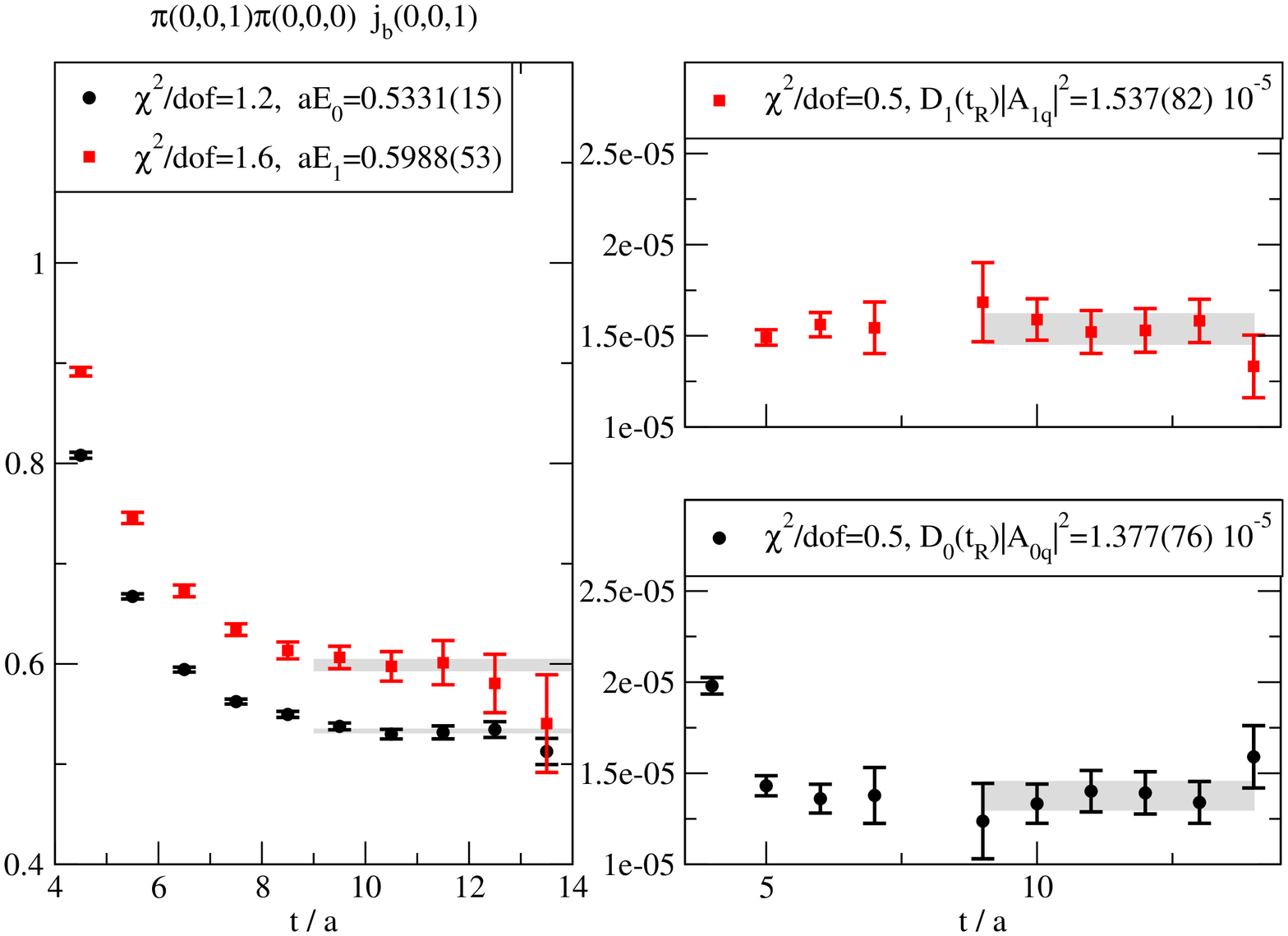}
  \caption{
    Same as Fig.~\ref{fig:mode1_mpi380}, but
    for the operator set \textcircled{2} and $m_\pi$ = 380~MeV.
  }
  \label{fig:mode2_mpi380}
\end{figure}

\begin{figure}[p]
  \centering
  \includegraphics[width=300pt,angle=270]
  {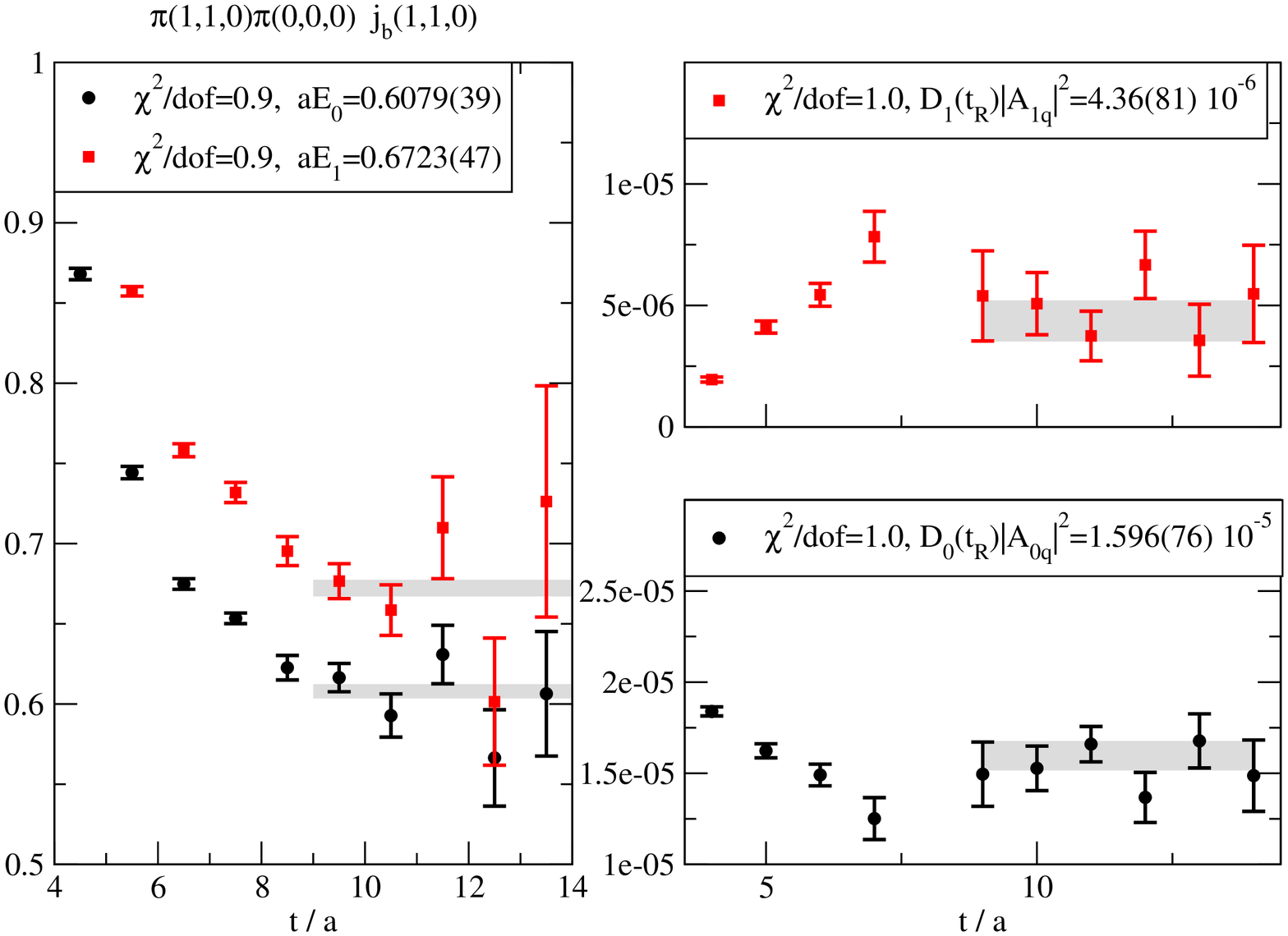}
  \caption{
    Same as Fig.~\ref{fig:mode1_mpi380}, but
    for the operator set \textcircled{3} and $m_\pi$ = 380~MeV.
  }
  \label{fig:mode3_mpi380}
\end{figure}

\begin{figure}[p]
  \centering
  \includegraphics[width=300pt,angle=270]
  {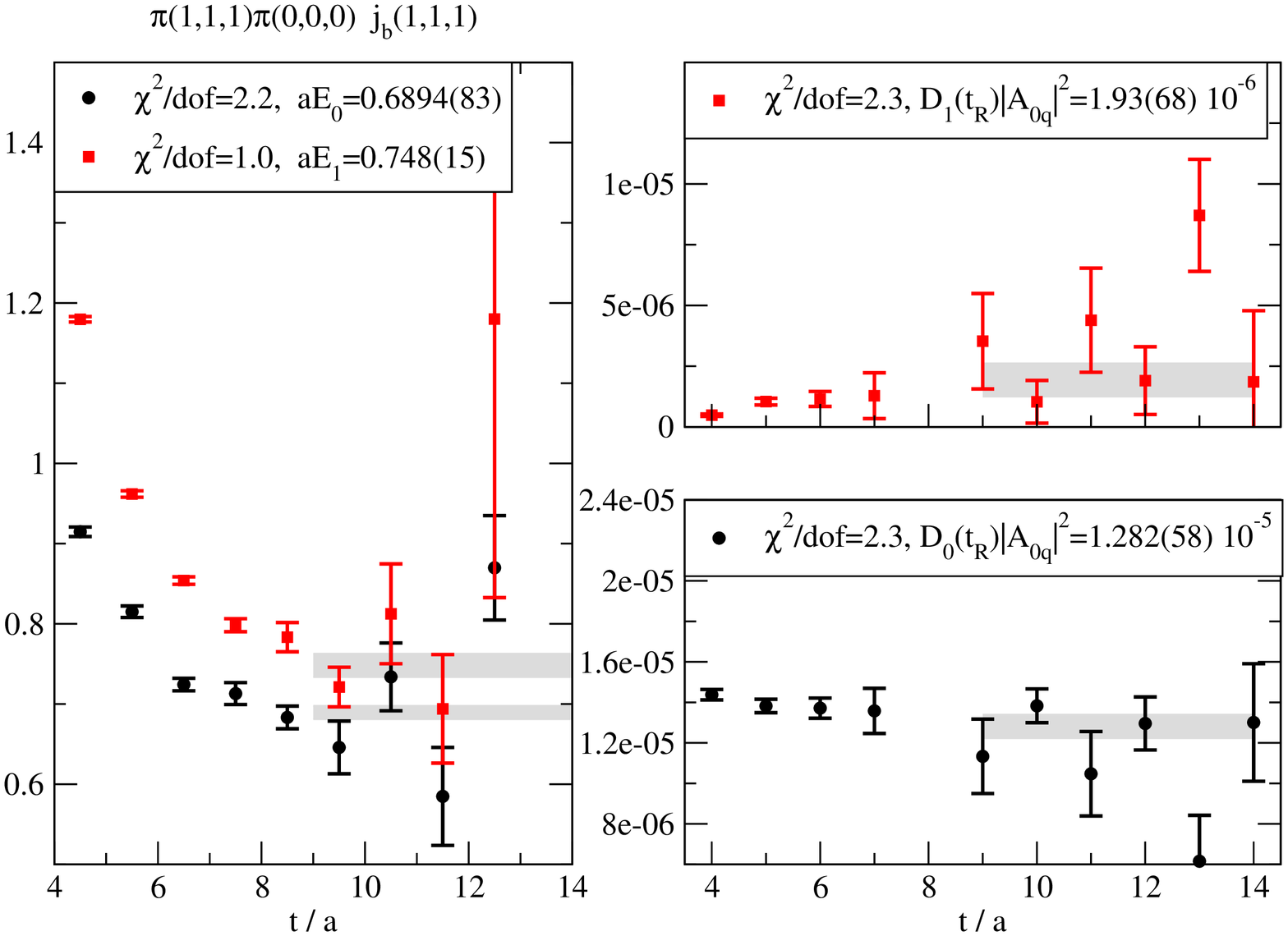}
  \caption{
    Same as Fig.~\ref{fig:mode1_mpi380}, but
    for the operator set \textcircled{4} and $m_\pi$ = 380~MeV.
  }
  \label{fig:mode4_mpi380}
\end{figure}

\begin{figure}[p]
  \centering
  \includegraphics[width=300pt,angle=270]
  {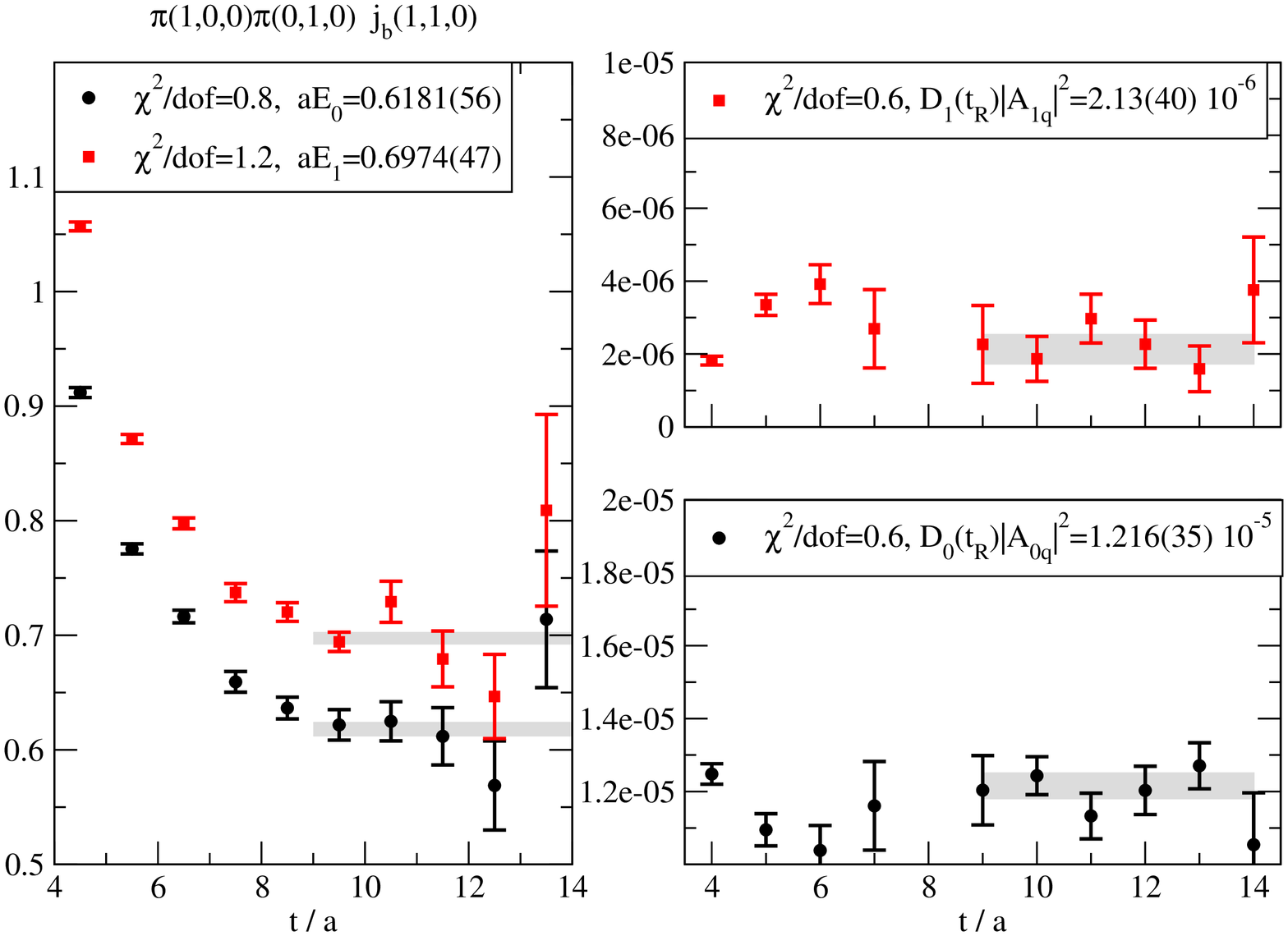}
  \caption{
    Same as Fig.~\ref{fig:mode1_mpi380}, but
    for the operator set \textcircled{5} and $m_\pi$ = 380~MeV.
  }
  \label{fig:mode5_mpi380}
\end{figure}

\begin{figure}[p]
  \centering
  \includegraphics[width=300pt,angle=270]
  {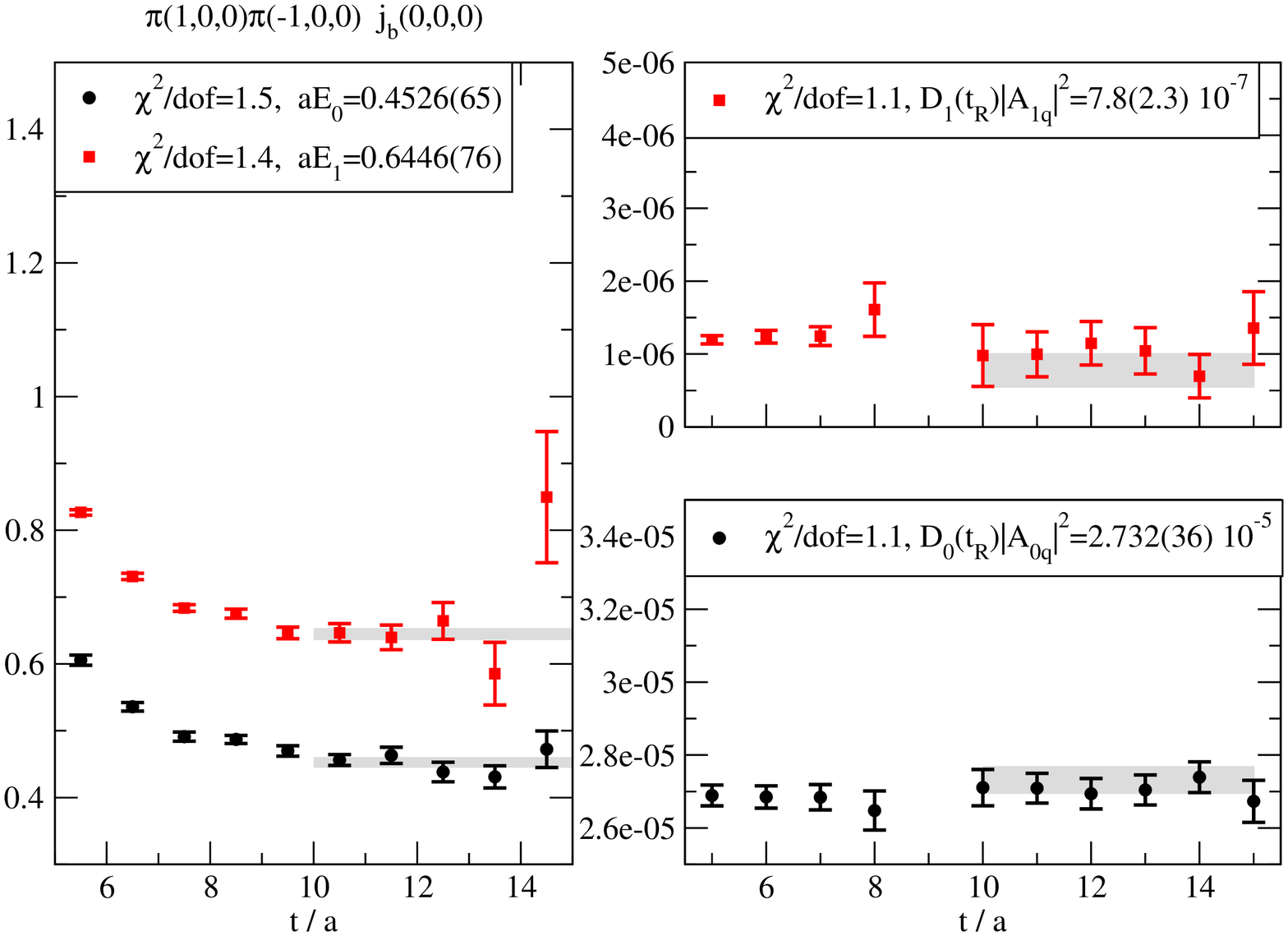}
  \caption{
    Same as Fig.~\ref{fig:mode1_mpi380}, but
    for the operator set \textcircled{1} and $m_\pi$ = 290~MeV.
  }
  \label{fig:mode1_mpi290}
\end{figure}

\begin{figure}[p]
  \centering
  \includegraphics[width=300pt,angle=270]
  {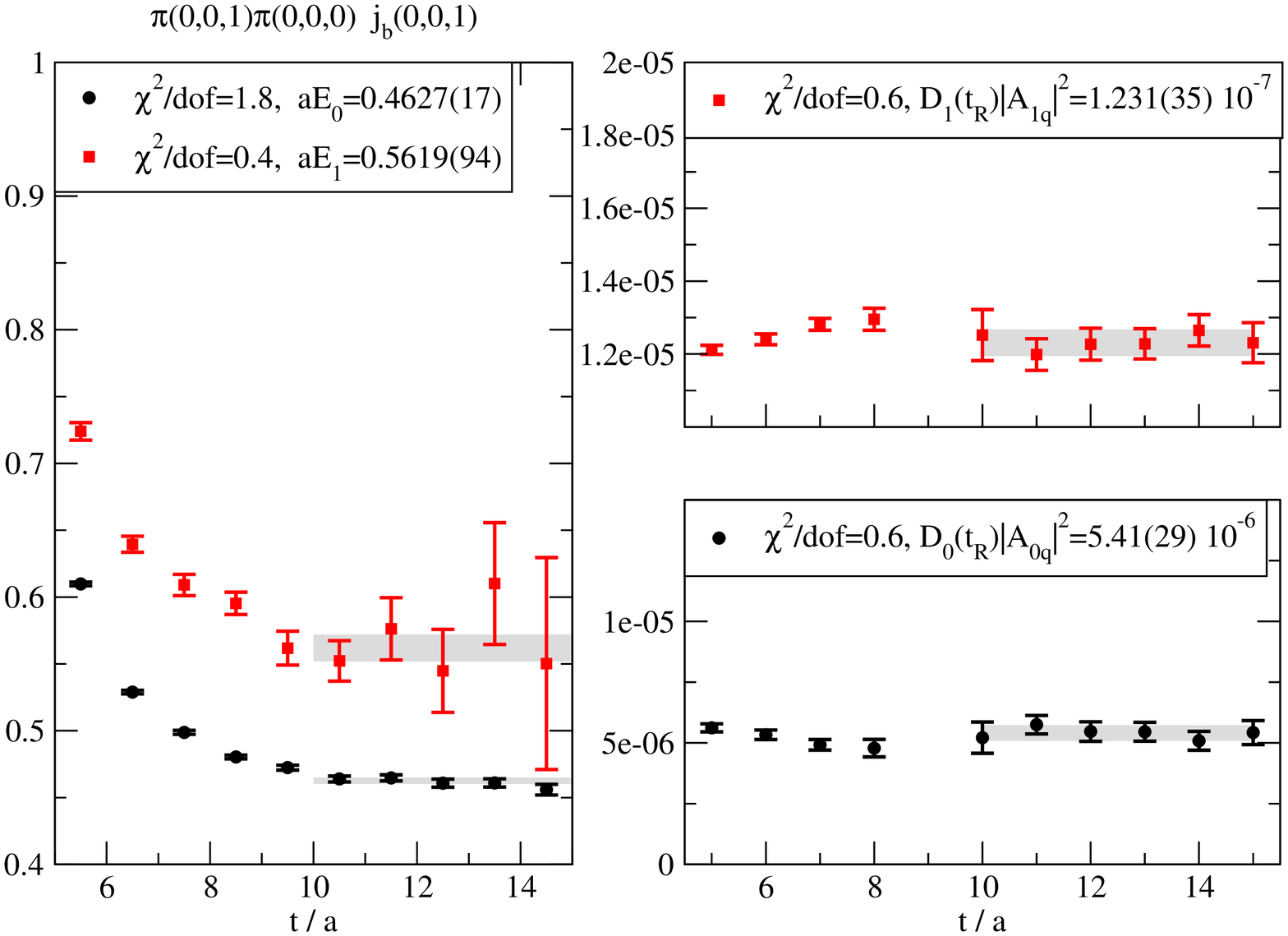}
  \caption{
    Same as Fig.~\ref{fig:mode1_mpi380}, but
    for the operator set \textcircled{2} and $m_\pi$ = 290~MeV.
  }
  \label{fig:mode2_mpi290}
\end{figure}

\begin{figure}[p]
  \centering
  \includegraphics[width=300pt,angle=270]
  {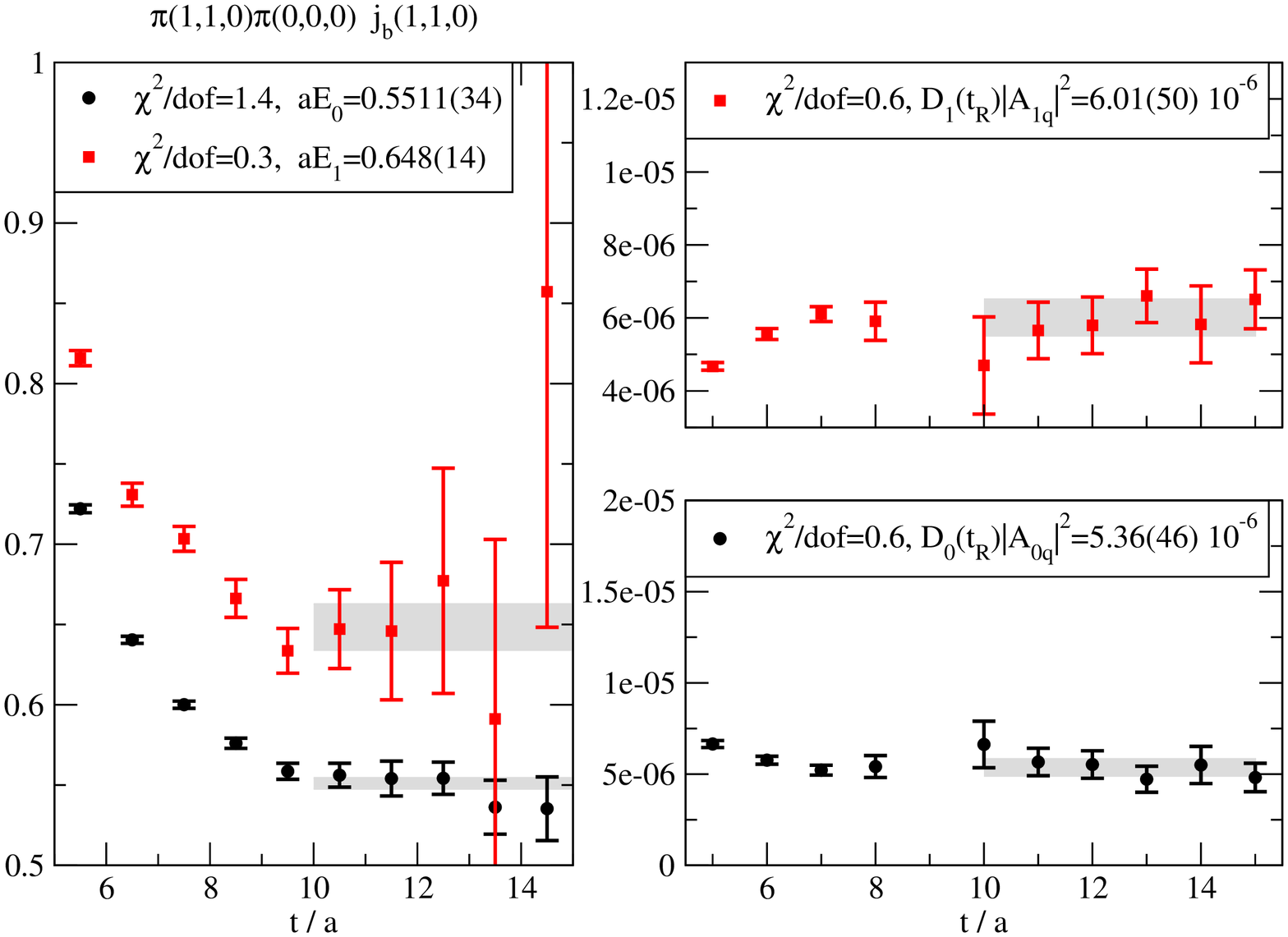}
  \caption{
    Same as Fig.~\ref{fig:mode1_mpi380}, but
    for the operator set \textcircled{3} and $m_\pi$ = 290~MeV.
  }
  \label{fig:mode3_mpi290}
\end{figure}

\begin{figure}[p]
  \centering
  \includegraphics[width=300pt,angle=270]
  {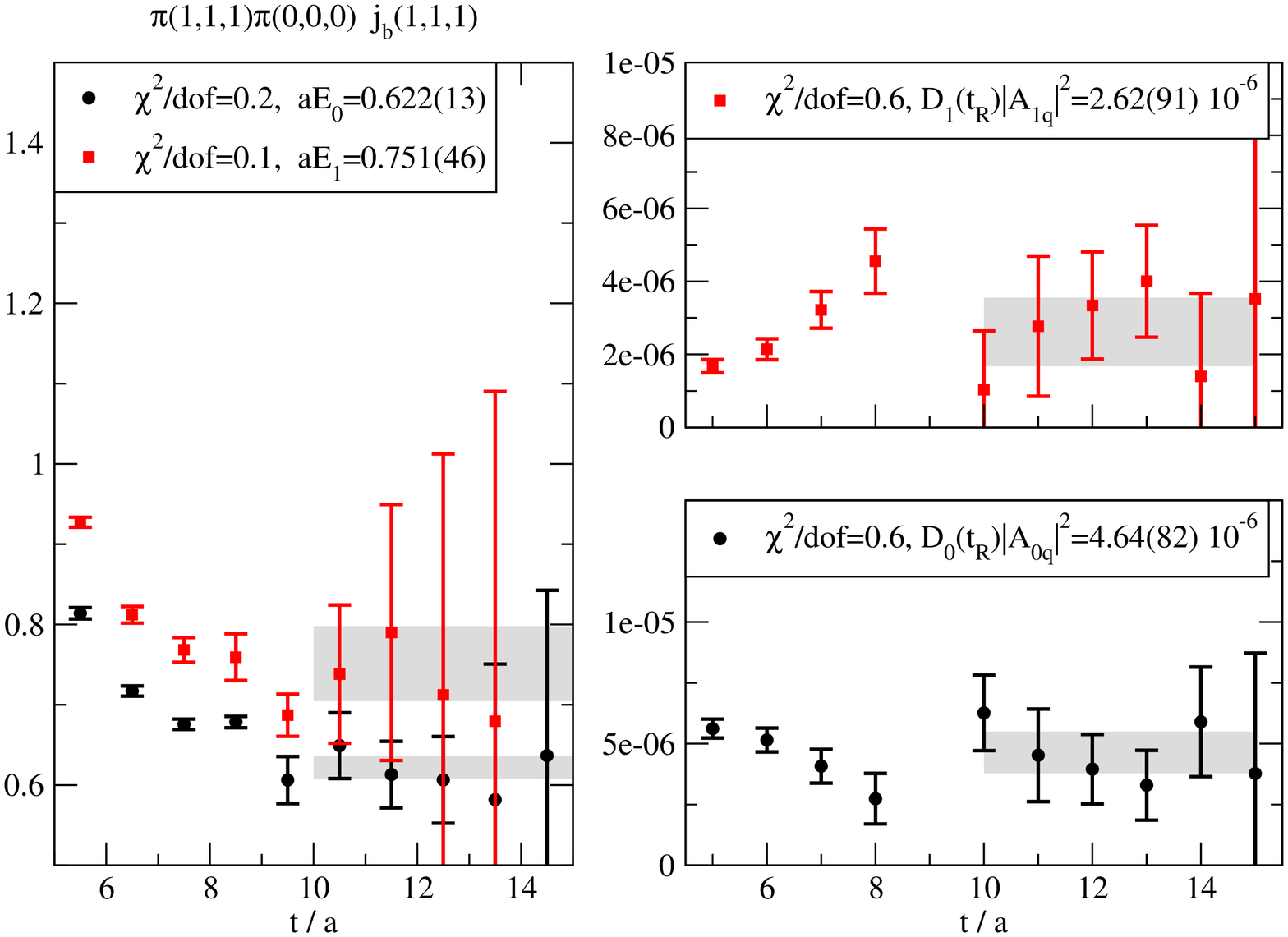}
  \caption{
    Same as Fig.~\ref{fig:mode1_mpi380}, but
    for the operator set \textcircled{4} and $m_\pi$ = 290~MeV.
  }
  \label{fig:mode4_mpi290}
\end{figure}

\begin{figure}[htbp]
  \centering
  \includegraphics[width=300pt,angle=270]
  {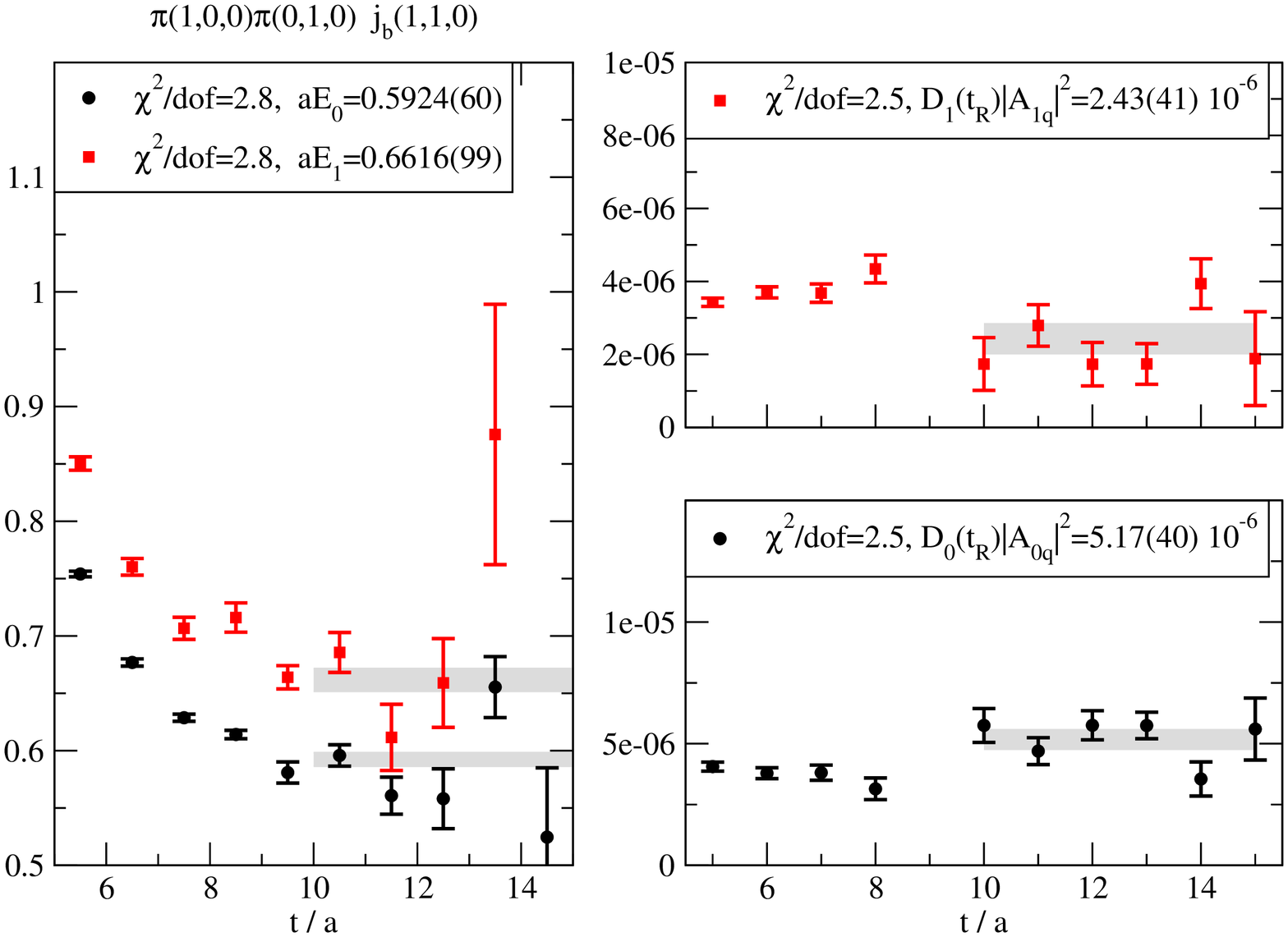}
  \caption{
    Same as Fig.~\ref{fig:mode1_mpi380}, but
    for the operator set \textcircled{5} and $m_\pi$ = 290~MeV.
  }
  \label{fig:mode5_mpi290}
\end{figure}

In practice, with a given reference time $t_R$, we determine $E_n$
by fitting the data of $D_n(t,t_R)$ to (\ref{eq:Dn_def}) and obtain
$D_n(t_R)|A_{n,q}|^2$ ($q=\bar{\psi}\psi$)
from (\ref{eq:amplitude}).
A fitting window of $t\in[t_R+a,t_R+6a]$ is used in our analysis. 
We gradually increase $t_R$ until the values of
$\chi^2/\rm{d.o.f}$ in the correlated fits are under control.
Here $\chi^2/\rm{d.o.f}$
is not a unique criterion to determine the fitting window. 
We also check the $t_R$ dependence to make sure that the effective mass
does not have systematically decreasing behavior. 
Also, given a pion mass, 
we try to have a consistent $t_R$ for different types of correlators, since
they have the same vector channel spectral weight function and the excited
states will have similar effects on the correlators. $t_R$ is chosen in a conservative 
way even at which $\chi^2/\rm{d.o.f}$ does not take its minimal value.
In this way, we set $t_R/a=8$ for $m_\pi$ = 380~MeV and $t_R/a=9$ for
$m_\pi$ = 290~MeV. 
The fit results are shown in
Figs.~\ref{fig:mode1_mpi380}--\ref{fig:mode5_mpi290} 
for each mass and the operator choices
\textcircled{1}, ..., \textcircled{5}.
In the left panel, the effective masses for the two lowest-energy
states are shown together with the fit results (gray bands).
We fix $t_R/a=$ 8 or 9. The effective mass at $t+a/2$ means 
an energy obtained from the equation that $ D_n(t+a)/D_n(t) = D_n(t+a,t_R)/D_n(t,t_R)$.
The right panel represents the effective amplitude
$D_n(t_R)|A_{n,q}|^2$ as a function of $t$.
The gray bands show the fitted value and the fitting range.
At the $t=t_R$, the data point for the amplitude is missing because
 $R(t,t_R)$ defined in Eq.~(\ref{eq:R_ratio}) is a unit matrix and thus contains no information for the amplitude. 
Although the signal quality depends on the mass and channel, energy
eigenstates are clearly identified for all channels.

\subsection{Results}

\begin{table}
  \centering
  \begin{tabular}{c@{\hskip 0.3in}rrr@{\hskip 0.3in}rrr}
    \hline
    & \multicolumn{3}{c@{\hskip 0.3in}}{$m_\pi=380$ MeV} &  
      \multicolumn{3}{c}{$m_\pi=290$ MeV} \\
    \hline
    No. & 
    \multicolumn{1}{c}{$E_n^*$} & 
    \multicolumn{1}{c}{$\delta_1$ ($^\circ$)} & 
    \multicolumn{1}{c@{\hskip 0.3in}}{$|F_\pi(s)|$} & 
    \multicolumn{1}{c}{$E_n^*$} & 
    \multicolumn{1}{c}{$\delta_1$ ($^\circ$)} &
    \multicolumn{1}{c}{$|F_\pi(s)|$}  \\
    \hline\hline
    \multirow{2}{*}{\textcircled{1}} & 
    876(7)   & 133.6(2.8) & 41.0(5.7)     & 
    796(12)   & 111.9(3.9) & 14.8(1.9)\\
    & 
    1203(8)  & 174.1(3.9) & 1.64(.14)     & 
    1134(13)  & 157.8(7.0) & 1.60(.26)\\
    \hline
    \multirow{2}{*}{\textcircled{2}} & 
    817(3)   & 4.95(.10)  & 9.28(.42)     & 
    671(4)    & 3.16(.25)  & 3.65(.14)\\
    & 
    947(10)  & 158.1(3.0) & 7.23(.29)     & 
    875(19)   & 140.1(5.2) & 7.36(.85)\\
    \hline
    \multirow{2}{*}{\textcircled{3}} & 
    848(9)   & 15.73(.87) & 19.9(4.0)     & 
    718(8)    & 8.3(1.1)   & 5.35(.34)\\
    & 
    987(10)  & 163.1(2.8) & 3.95(.35)     & 
    936(31)   & 139.3(8.2) & 4.83(.21)\\
    \hline
    \multirow{2}{*}{\textcircled{4}} & 
    913(19)  & 18.9(5.4)  & 13.0(4.2)     & 
    750(34)   & 14.3(6.5)  & 7.6(1.9)\\
    & 
    1047(32) & 152(23)   & 4.2(3.2)      & 
    1054(101) & 133(31)    & 3.78(.62)\\
    \hline
    \multirow{2}{*}{\textcircled{5}} & 
    871(12)  & 52.7(5.6)  & 41.6(5.9)     & 
    813(13)   & 21.8(5.0)  & 14.3(1.3)\\
    & 
    1040(10) & 164.9(3.5) & 3.26(.29)     & 
    964(21)   & 150.1(6.8) & 3.73(.31)\\
    \hline
  \end{tabular}
  \caption{
    Center-of-mass energy $E_n^*$, $P$-wave pion-pion scattering phase
    shift $\delta_1$ and the modulus of the pion form factor at the
    pion masses $m_\pi$ = 380~MeV (left block) and 290~MeV (right).
    $E_n^*$ are given in units of MeV.
  }
  \label{tab:E_and_A}
\end{table}

We convert the energy eigenvalues $E_n$ ($n$ = 0 and 1) for each
operator choice, {\it i.e.} the momentum configuration,
into the center-of-mass energy $E_n^*$ using the dispersion relation. 
Then, inserting $E_n^*$ into the L\"uscher's formula
(\ref{eq:Luscher_formula}) yields the $P$-wave scattering phase shift
$\delta_1$.
The results for $E_n^*$ and $\delta_1$ are shown in 
Table~\ref{tab:E_and_A}. 
We neglect the $K\bar{K}$ multichannel effects since the largest energy
$E_n^*$ listed in Table~\ref{tab:E_and_A} is only slightly higher
than $2m_K$.

\begin{figure}[tbp]
  \centering
  \includegraphics[width=340pt,angle=270]{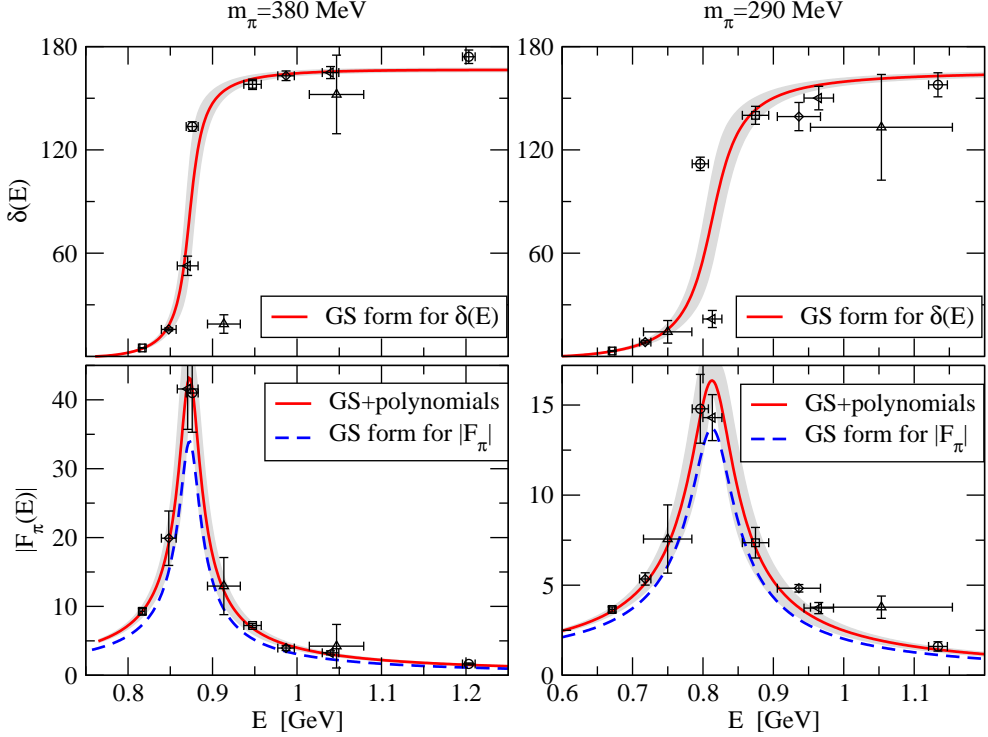}
  \caption{
    Upper panels:
    Scattering phases calculated using the L\"uscher 
    formula~(\ref{eq:Luscher_formula})
    together with the fits to the GS form~(\ref{eq:phase_shift}).
    Lower panels:
    Modulus of the pion form factor calculated using the
    Lellouch-L\"uscher formula~(\ref{eq:LL_formula}) together with 
    the GS-model curves (blue dashed) and the fits to (\ref{eq:ff})
    (red solid).
    Circles, squares, diamonds, triangles-up and triangles-left data
    points correspond to the operator sets
    \textcircled{1}--\textcircled{5} given in
    Table~\ref{tab:operator}, respectively. 
  }
  \label{fig:form_factor}
\end{figure}

In the upper panels of Fig.~\ref{fig:form_factor} we plot the
scattering phase $\delta_1$ at various energies $E_n^*$. 
To study the energy dependence of $\delta_1$, we fit the lattice
data to the GS model~(\ref{eq:phase_shift}). 
We find that this model gives a rather good description of the lattice
data. 
Through the fit, we can extract the $g_{\rho\pi\pi}$ coupling and the
$\rho$-resonance mass $m_\rho$, 
which are listed in Table~\ref{tab:parameter}. 
Such way to determine the $\rho$-resonance mass is different from the
conventional method to obtain the effective mass from a two-point correlation function.
We can make a comparison of $m_\rho$ given in Table~\ref{tab:parameter} and the
effective mass of operator choice $\textcircled{1}$ given in Table~\ref{tab:E_and_A}.
As the pion mass decreases, the effective mass becomes smaller than the
$m_\rho$ extracted from the scattering phase. This is consistent with our expectation,
since at the physical pion mass, the effective mass of the ground $\pi\pi$ state shall be 
significantly lower than the physical $\rho$-meson mass. To see this
trend more clearly, we still need to improve precision or to use lighter
pion mass.

Near the resonance region, some data points deviate from the fit curve
significantly.
This might be due to the rapid change of the scattering phase in
the resonance region.
Namely, some systematic effects in the determination of the energy
eigenvalues may translate into a big shift in the scattering phase and
cause a deviation from the fit curve. 
For instance, in our calculation we use only $2\times2$ correlation
matrix, which might not be enough to completely eliminate the
excited-state effects.

With the values of $|A_{n,q}|$, we determine the modulus of the pion
form factor $|F_\pi(s)|$ using the Lellouch-L\"uscher formula (\ref{eq:LL_formula}). 
In this formula, a derivative of scattering phase is required. 
Here we use the GS description of the scattering phase
(\ref{eq:phase_shift}). 
The results for $|F_\pi(s)|$ are given in Table~\ref{tab:E_and_A}. 
In the lower panels of Fig.~\ref{fig:form_factor}, $|F_\pi(s)|$ is
shown as a function of energy. 
As mentioned before, the simple GS form (\ref{eq:simple_GS}) 
(using the lattice results of $m_\rho$ and $g_{\rho\pi\pi}$ in
Table~\ref{tab:parameter} as inputs) shown by the dashed curve
gives too small values near the resonance region
compared to our lattice data.

\begin{figure}[tbp]
  \centering
  \includegraphics[width=340pt,angle=270]{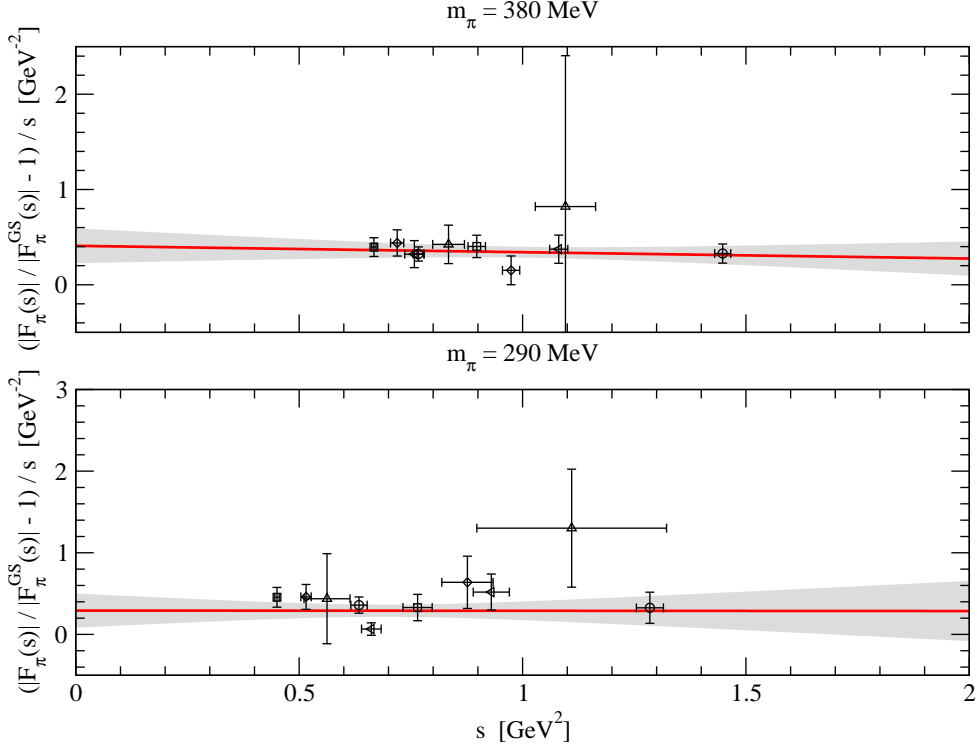}
  \caption{
    Difference between the lattice data of $|F_\pi(s)|$ and the GS
    form~(\ref{eq:simple_GS}).
    The data for $(|F_\pi(s)/F_\pi^{GS}(s)|-1)/s$ are 
    plotted as a function of $s$ together with the fit to the 
    polynomial $c_1+c_2(s-2m_\rho^2)$.
  }
  \label{fig:form_factor_diff}
\end{figure}

\begin{table}
  \centering
  \begin{tabular}{ccc@{\hskip 0.3in}ccc}
    \hline
    \multicolumn{3}{c@{\hskip 0.3in}}{$m_\pi=380$ MeV} &
    \multicolumn{3}{c}{$m_\pi=290$ MeV} \\
    \hline
    \multicolumn{1}{c}{$c_0$} &  
    \multicolumn{1}{c}{$c_1$} & 
    \multicolumn{1}{c@{\hskip 0.3in}}{$c_2$} & 
    \multicolumn{1}{c}{$c_0$} &  
    \multicolumn{1}{c}{$c_1$} &  
    \multicolumn{1}{c@{\hskip 0.3in}}{$c_2$} \\
    \hline\hline
    1.273(51) & 0.31(10) & $-0.07(17)$  &       
    1.195(47) & 0.29(19) & $-0.00(27)$\\
    \hline
  \end{tabular}
  \caption{
    Coefficients $c_0$, $c_1$ and $c_2$ of the model (\ref{eq:ff}).
    $c_1$ and $c_2$ are determined by fitting the lattice data of 
    $(|F_\pi(s)/F_\pi^{GS}(s)|-1)/s$ to the polynomials 
    $c_1+c_2(s-2m_\rho^2)$
    and $c_0$ is determined by charge conservation condition: 
    $c_0+c_1(-m_\rho^2)+c_2(-m_\rho^2)^2=1$.
    $c_1$ and $c_2$ are given in units of GeV$^{-2}$ and GeV$^{-4}$,
    respectively.
  }
  \label{tab:c_n}
\end{table}

We then use the modified form (\ref{eq:ff}) to describe the lattice
data.  
The difference between the form~(\ref{eq:simple_GS}) and (\ref{eq:ff})
can be written as
\begin{equation}
  \frac{|F_\pi(s)|}{|F_\pi^{GS}(s)|}-1 = 
  \sum_{n=0}^N
  c_n \left( (s-m_\rho^2)^n-(-m_\rho^2)^n \right)
  =
  s\left( c_1+c_2(s-2m_\rho^2)+\cdots \right).
\end{equation}
In Fig.~\ref{fig:form_factor_diff} we show the data of
$(|F_\pi(s)/F_\pi^{GS}(s)|-1)/s$ as a function of $s$. 
The data points seem to be well described by a straight line up to
statistical fluctuations.
We therefore fit them to the form $c_1+c_2(s-2m_\rho^2)$. 
The fitting results for $c_1$ and $c_2$, together with $c_0$
determined from charge conservation, are given in Table~\ref{tab:c_n}.
Within current statistics, the values of $c_2$ are consistent with $0$
for both pion masses,
and it is not necessary to pursue higher polynomial terms with
$c_{n>2}$.
Putting $c_0$, $c_1$ and $c_2$ into (\ref{eq:ff}), we draw the fit
curves for $|F_\pi(s)|$ in Fig.~\ref{fig:form_factor}. 
By including the polynomial terms, the curves match the lattice data.
Note that we have imposed the charge conservation condition
when obtaining the values of $c_n$ in Table~\ref{tab:c_n}. If we do not impose
this constraint and fit with a free $c_0$, we find for 
$c_0+c_1(-m_\rho^2)+c_2(-m_\rho^2)^2=1.08(14)$ at $m_\pi=$ 380 MeV and
1.12(16) at $m_\pi=290$ MeV. The charge conservation condition is well reproduced
by our lattice data.

As a by-product of this calculation, we evaluate the pion mean-square
charge radius (isovector part only) through
\begin{equation}
  \label{eq:charge_radius}
  \langle r_\pi^2\rangle = 
  6\frac{\partial |F_\pi(s)|}{\partial s}\bigg|_{s=0}
  = 6\left(
    -\frac{1}{f_0}\left(\frac{b}{4}+\frac{1}{3\pi}\right)
    +c_1+c_2\left(-2m_\rho^2\right)
  \right),
\end{equation}
using the modified GS form.
The first term arises from the GS model with $b$ and $f_0$ defined in (\ref{eq:b_and_f0}).
The second and third terms are the polynomial corrections.
The results for $\langle r_\pi^2\rangle$ are listed in Table~\ref{tab:parameter}, where 
they are compared with the calculation in the spacelike
momentum transfer on the same gauge ensembles \cite{Kaneko:2010ru,Fukaya:2014jka}.
The central values of the timelike data seem systematically larger than
the spacelike ones but still consistent within the statistical errors.

\begin{table}
  \centering
  \begin{tabular}{c|c|c}
    \hline
    Lattice & $m_\pi$ = ``380~MeV'' & $m_\pi$ = ``290~MeV''\\
    \hline\hline
    $m_\pi$ (MeV) &  $378.6(7)$ & $291.8(1.1)$ 
    \\
    $m_\rho$ (MeV) & $875(7)$  & $819(14)$ 
    \\
    $g_{\rho\pi\pi}$ & 5.85(19)  & 5.78(23) 
    \\
    (timelike) $\langle r_\pi^2\rangle$ (fm$^2$) & 
    0.377(38) & 0.392(41) 
    \\
    (spacelike) $\langle r_\pi^2\rangle$ (fm$^2$) & 
    0.334(10)($^{+00}_{-32}$) & 0.366(19)($^{+00}_{-42}$)
    \\
    \hline
  \end{tabular}
  \caption{
    Numerical results for $m_\pi$, $m_\rho$, $g_{\rho\pi\pi}$ and 
    $\langle r_\pi^2\rangle$ at $m_\pi$ = 380~MeV (left) and 290~MeV
    (right). The timelike $\langle r_\pi^2\rangle$ are evaluated
    using Eq.~(\ref{eq:charge_radius}). The spacelike $\langle r_\pi^2\rangle$
    are compiled using the spacelike form factor, where the first
    error is statistical and the second one originates from the choice of 
    the parametrization form of the $q^2$ dependence of $F_\pi(q^2)$ 
    (linear, quadratic, VMD with polynomial corrections).}
  \label{tab:parameter}
\end{table}

\section{Conclusion}
In this work, we calculate the complex phase and the modulus of the
pion form factor in the timelike momentum region.
We perform the calculation at two pion masses $m_\pi$ = 380~MeV and
290~MeV and at a lattice spacing of $a$ = 0.11~fm on
$N_f=2+1$-flavor overlap fermion configurations generated by the JLQCD
Collaboration. 

In the elastic scattering region, the complex phase of $F_\pi(s)$ is
given by the $P$-wave pion-pion scattering phase, and thus can be evaluated
using the standard L\"uscher's finite-volume formula. 
We obtain the results at ten different values of $s$ from one setup in
the center-of-mass frame and four in the moving frames.
From the energy dependence of the scattering phase, we extract the
$g_{\rho\pi\pi}$ coupling constant and the $\rho$-resonance mass
$m_\rho$.

Lattice calculation of the modulus of the pion form factor was
originally proposed in \cite{Meyer:2011um}, and here we extend the
method to general moving frames and perform the actual calculation
using the all-to-all propagator technique.
We obtain a clear signal of the form factor and phase indicating the
vector meson resonance.
The lattice data for $|F_\pi(s)|$ are not consistent with the simple
GS model. 
To address this discrepancy we introduce a simple polynomial
correction to the GS form, which describes the lattice data quite well.

Though we focus on the calculation of the matrix elements 
$\langle0|j_{\bf b}^{\bar{\psi}\psi}|\pi\pi\rangle_V$, 
which can be directly related to $|F_\pi(s)|$, 
the information hidden in the matrix elements of the 
$j_{\bf b}^{\pi\pi}$-current insertion can also be
useful for the study of the resonance properties 
\cite{Liu:2008hy,Niu:2009gt,Niu:2010rk}.

As an exploratory study, our work demonstrates the feasibility of
calculating the pion form factor in the timelike region using lattice
QCD. 
It is still challenging to make a precise comparison to the
experimental $e^+e^-$ data, since we need to calculate the form factor
at the physical pion mass, extract many more data points and control
the errors both statistically and systematically at the level of
experimental precision.

\begin{acknowledgments}
We thank our JLQCD colleagues for many valuable suggestions 
and encouragement.
X.F. would like to thank Professor Norman H. Christ for very helpful
discussions. 
Numerical simulations are performed on Hitachi SR16000 at the High Energy
Accelerator Research Organization under the support of its Large Scale
Simulation Program (No. 12/13-04 and 13/14-04), as well as on another
Hitachi SR16000 at the
Yukawa Institute for Theoretical Physics, Kyoto University.
This work is supported in part by the Grant-in-Aid of the Japanese
Ministry of Education (Grants No. 21674002, No. 25287046, No. 26247043, 
and No. 26400259), 
by MEXT SPIRE and JICFuS and by U.S. DOE Grant No. DE-SC0011941.
\end{acknowledgments}

\appendix
\section{Gounaris-Sakurai Model}
\label{sec:GS_model}
Using the twice-subtracted dispersion relation, one can relate the real
part of $\Pi_\rho(s)$ to its imaginary part through
\begin{equation}
  {\rm Re}\,\Pi_\rho(s) =
  c_0+c_1s+\frac{s^2}{\pi}{\mathcal P}\int_{4m_\pi^2}^\infty
  ds'\,\frac{{\rm Im}\,\Pi_\rho(s')}{{s'}^2(s'-s)},
\end{equation}
where ${\mathcal P}$ denotes the principal value of the integral.
Inserting (\ref{eq:im_Pi}) into the dispersion relation, one has
\begin{equation}
  {\rm Re}\,\Pi_\rho(s) =  
  c_0+c_1s+ \frac{g_{\rho\pi\pi}^2}{6\pi} 
  \left( 
    k^2h(\sqrt{s})-\frac{s}{3\pi}+\frac{m_\pi^2}{\pi}
  \right),
\end{equation}
where the function $h(\sqrt{s})$ is given by
\begin{equation}
  h(\sqrt{s}) = \frac{2}{\pi} \frac{k}{\sqrt{s}} \ln\left(
    \frac{\sqrt{s}+2k}{2m_\pi}
  \right),
\end{equation}
for $s>4m_\pi^2$.
Using the conditions
\begin{equation}
  {\rm Re}\,\Pi_\rho(s)\bigg|_{s=m_\rho^2}=0,
  \quad 
  \frac{d\,{\rm Re}\,\Pi_\rho(s)}{ds}\bigg|_{s=m_\rho^2}=0,
\end{equation}
one can determine the constants $c_0$ and $c_1$ and find for
\begin{equation}
  {\rm Re}\,\Pi_\rho(s) = \frac{g_{\rho\pi\pi}^2}{6\pi}
  \left(
    k^2(h(\sqrt{s})-h(m_\rho))-\frac{2k_\rho^2}{m_\rho}h'(m_\rho)(k^2-k_\rho^2)
  \right).
\end{equation}
This finally results in the GS form factor as
\begin{equation}
  \label{eq:GS_FF}
  F_\pi^{GS}(s) = \frac{f_0}{
    k^2h(\sqrt{s})-k_\rho^2h(m_\rho)+b(k^2-k_\rho^2)-\frac{k^3}{\sqrt{s}}i
  }
\end{equation}
with
\begin{eqnarray}
  \label{eq:b_and_f0}
  b & = & 
  -h(m_\rho) -\frac{24\pi}{g^2_{\rho\pi\pi}}
  -\frac{2k_\rho^2}{m_\rho}h'(m_\rho),
  \nonumber\\
  f_0 & = &
  -\frac{m_\pi^2}{\pi}-k_\rho^2h(m_\rho)-b\frac{m_\rho^2}{4}.
\end{eqnarray}
Here we use the same notations as in \cite{Francis:2013fzp}.

Using the Watson's theorem, it is natural to find for the $P$-wave
pion-pion scattering phase
\begin{equation}
  \label{eq:phase_shift}
  \frac{k^3}{\sqrt{s}}\cot\delta_1(s) = 
  k^2h(\sqrt{s})-k_\rho^2h(m_\rho)+b(k^2-k_\rho^2).
\end{equation}
Near the resonance energy $\sqrt{s}\sim m_\rho$, one has
\begin{equation}
  \frac{k^3}{\sqrt{s}}\cot\delta_1(s) = 
  -\frac{24\pi}{g_{\rho\pi\pi}^2}(k^2-k_\rho^2)+O((\sqrt{s}-m_\rho)^2).
\end{equation} 
This approximation reproduces the effective range formula, which was
proposed in \cite{Luscher:1991cf} and commonly used in previous
lattice QCD studies
\cite{Aoki:2007rd,Gockeler:2008kc,Feng:2010es,Lang:2011mn,Aoki:2011yj,Pelissier:2012pi,Dudek:2012xn} 
to describe the $s$ dependence of the scattering phase.
Note that both the GS model and effective range formula account for the leading-order Taylor expansion term at $\sqrt{s}=m_\rho$ and
thus have no control of the $s$ dependence for $\sqrt{s}\gg m_\rho$. 
In \cite{Dudek:2012xn}, various barriers were set for large $s$ but
with the given statistics different parametrizations are not
distinguishable.
Considering the fact that the current calculation mainly collects the
data near the resonance energy, we simply adopt (\ref{eq:phase_shift})
in our analysis.

\section{L\"uscher's formula used in this calculation}
\label{sect:Luscher_formula}
Given the total momentum ${\bf P}$ and irreducible representation
$\Gamma$,
the ways to construct the function $\phi^{{\bf P},\Gamma}(q)$ are
given in \cite{Luscher:1990ux} for the center-of-mass frame
and in \cite{Rummukainen:1995vs} for the general moving frames.
Here we simply give the expressions for $\phi^{{\bf P},\Gamma}(q)$,
which are defined through
\begin{equation}
  \tan \phi^{{\bf P},\Gamma}(q) =
  -\frac{\gamma\pi^{3/2}q}{Z^{{\bf d},\Gamma}(q)},
  \quad 
  {\bf P}=\frac{2\pi}{L}{\bf d}
\end{equation}
with no ambiguity by setting $\phi^{{\bf P},\Gamma}(0)=0$ and
requiring a continuous dependence of $\phi^{{\bf P},\Gamma}(q)$ on $q$.
The denominator $Z^{{\bf d},\Gamma}(q)$ is given by
\begin{equation}
  \label{eq:phi_expression}
  \begin{array}{ll}
    {\mathcal Z}_{00}^{\bf d},
    \quad & \text{for } 
    {\bf d}=(0,0,0),\, \Gamma=T_1^-,
    \\
    {\mathcal Z}_{00}^{\bf d} 
    +\frac{2}{\sqrt{5}}q^{-2}{\mathcal Z}_{20}^{\bf d},
    \quad & \text{for } 
    {\bf d}=(0,0,1),\, \Gamma=A_2^-,
    \\
    {\mathcal Z}_{00}^{\bf d}
    -\frac{1}{\sqrt{5}}q^{-2}{\mathcal Z}_{20}^{\bf d}
    +i\frac{\sqrt{3}}{\sqrt{10}}q^{-2}
    ({\mathcal Z}_{22}^{\bf d}-{\mathcal Z}_{2\bar{2}}^{\bf d}),
    \quad & \text{for }
    {\bf d}=(1,1,0),\, \Gamma=B_1^-,
    \\
    {\mathcal Z}_{00}^{\bf d}
    -\frac{1}{\sqrt{5}}q^{-2}{\mathcal Z}_{20}^{\bf d}
    -i\frac{\sqrt{3}}{\sqrt{10}}q^{-2}
    ({\mathcal Z}_{22}^{\bf d} -{\mathcal Z}_{2\bar{2}}^{\bf d}), 
    \quad &\text{for }
    {\bf d}=(1,1,0),\, \Gamma=B_2^-,
    \\
    {\mathcal Z}_{00}^{\bf d}
    +\frac{\sqrt{2}}{\sqrt{15}}q^{-2}
    \left(
      (-1-i){\mathcal Z}_{21}^{\bf d}
      +(1-i){\mathcal Z}_{2\bar{1}}^{\bf d}
      +i{\mathcal Z}_{22}^{\bf d}
      -i{\mathcal Z}_{2\bar{2}}^{\bf d}
    \right),
    \quad & \text{for } 
    {\bf d}=(1,1,1),\, \Gamma=A_2^-.
  \end{array}
\end{equation}
In the above expression, ${\mathcal Z}_{lm}^{\bf d}$ is a short-hand
notation for the zeta function
${\mathcal Z}_{lm}^{\bf d}(1;q^2)$, which is defined through
\begin{equation}
  {\mathcal Z}_{lm}^{\bf d}(s;q^2) = \sum_{{\bf n}\in P_{\bf d}}
  \frac{{\mathcal Y}_{lm}^*({\bf n})}{(|{\bf n}|^2-q^2)^s},
\end{equation}
with
\begin{equation}
  {\mathcal Y}_{lm}({\bf r})=r^lY_{l,m}(\Omega_{\bf r}),
  \quad 
  {\mathcal Y}_{l\bar{m}}({\bf r})=r^lY_{l,-m}(\Omega_{\bf r})
\end{equation}
and
\begin{equation}
  P_{\bf d} = 
  \left\{
    {\bf n}\,\bigg|\,{\bf n} =
    \vec{\gamma}^{-1}({\bf m}+\frac{1}{2}{\bf d}),
    \quad \text{for }
    {\bf m}\in {\mathbb Z}^3
  \right\}.
\end{equation}
${\mathcal Z}_{lm}^{\bf d}(s;q^2)$ is divergent for $s\le\frac{l}{2}+\frac{3}{2}$ and needs to be analytically continued
in a numerical calculation.
An analytically continued form of ${\mathcal Z}_{lm}^{\bf d}(1;q^2)$
is given in \cite{Feng:2011ah} and confirmed by \cite{Leskovec:2012gb}
with detailed derivations.\footnote{
  In \cite{Leskovec:2012gb}, the zeta function is defined using
  ${\mathcal Y}_{lm}({\bf n})$ rather than its complex conjugate.}

\section{Lellouch-L\"uscher formula in the $P$-wave $\pi\pi$ scattering}
\label{sect:LL_Pwave}
The demonstration of (\ref{eq:LL_formula}) follows closely
\cite{Lin:2001ek}.

In the infinite volume limit, the correlator $C_V(t)$ turns out to be
\begin{eqnarray}
  \label{eq:corr_infty_limit}
  C_V(t) & = & 
  \int_V d^3{\bf x}\; e^{-i{\bf P}\cdot{\bf x}}
  \langle0|j_{\bf b}({\bf x},t)j_{\bf b}^\dagger({\bf 0},0)|0\rangle
  \nonumber\\
  & \xrightarrow[V\rightarrow\infty]{} & 
  \frac{1}{(2\pi)^3} \int 
  \frac{d^3{\bf p}_1}{2E_1}\frac{d^3{\bf p}_2}{2E_2} \;
  \delta^{(3)}({\bf p}_1+{\bf p}_2-{\bf P})
  | \langle0|j_{\bf b}(0)|\pi\pi\rangle |^2
  e^{-(E_1+E_2)t}
  \nonumber\\
  & = &
  \frac{1}{(2\pi)^3}\int dE 
  \int \frac{d^3{\bf p}_1}{2E_1}\frac{d^3{\bf p}_2}{2E_2} \;
  \delta^{(3)}({\bf p}_1+{\bf p}_2-{\bf P})\delta(E-E_1-E_2)
  |\langle0|j_{\bf b}(0)|\pi\pi\rangle|^2 e^{-Et}.
  \nonumber\\
\end{eqnarray}
In a general moving frame, the center of mass is moving with velocity
${\bf v}={\bf P}/E$ 
and the momenta ${\bf p}_i$ and ${\bf p}_i^*$ (center-of-mass momentum) are related to each other by the standard Lorentz transformation
\begin{eqnarray}
  &&
  {\bf p}_1=\vec{\gamma}({\bf p}_1^*+{\bf v}E_1^*),
  \quad 
  {\bf p}_2=\vec{\gamma}({\bf p}_2^*+{\bf v}E_2^*)
  \nonumber\\
  && 
  E_1=\gamma(E_1^*+{\bf v}\cdot{\bf p}_1^*),
  \quad 
  E_2=\gamma(E_2^*+{\bf v}\cdot{\bf p}_2^*),
\end{eqnarray}
where we have defined
\begin{equation}
  \gamma=\frac{1}{\sqrt{1-{\bf v}^2}},
  \quad
  \vec{\gamma}{\bf p}=\gamma {\bf p}_\parallel+{\bf p}_\perp,
  \quad
  \vec{\gamma}^{-1}{\bf p}=\gamma^{-1} {\bf p}_\parallel+{\bf p}_\perp,
\end{equation}
with 
${\bf p}_\parallel=\frac{{\bf p}\cdot {\bf v}}{|{\bf v}|^2}{\bf v}$ 
and 
${\bf p}_\perp={\bf p}-{\bf p}_\parallel$.
Note that the measure $\frac{d^3{\bf p}_i}{2E}$ and delta function $\delta^{(4)}(p_1+p_2-P)$ are Lorentz invariant and satisfy
\begin{equation}
  \label{eq:lorentz1}
  \frac{d^3{\bf p}_i}{2E_i}=\frac{d^3{\bf p}_i^*}{2E_i^*},
  \quad
  \delta^{(4)}(p_1+p_2-P)=\delta^{(4)}(p_1^*+p_2^*-P^*),
  \quad 
  P^*=(E^*,{\bf 0}).
\end{equation}
However, the amplitude $\langle0|j_{\bf b}(0)|\pi\pi\rangle$ is not invariant and transforms as
\begin{eqnarray}
  \label{eq:lorentz2}
  \langle0|j_{\bf b}(0)|\pi\pi\rangle & = &
  i({\bf p}_1-{\bf p}_2)\cdot{\bf b}\;F_\pi(s)
  \nonumber\\
  & = & 
  i[\vec{\gamma}({\bf p}_1^*-{\bf p}_2^*)]\cdot{\bf b}\; F_\pi(s)
  \nonumber\\
  & = &
  ig(\gamma)({\bf p}_1^*-{\bf p}_2^*)\cdot{\bf b}\;F_\pi(s),
\end{eqnarray}
with $g(\gamma)=\gamma$ for 
${\bf b}\parallel{\bf P}$ and $g(\gamma)=1$ for 
${\bf b}\perp{\bf P}$.

Inserting (\ref{eq:lorentz1}) and (\ref{eq:lorentz2}) into (\ref{eq:corr_infty_limit}), we have
\begin{eqnarray}
  \label{eq:large_V1}
  C(t) &\xrightarrow[V\rightarrow\infty]{} &
  \frac{1}{(2\pi)^3} \int dE 
  \int \frac{d^3{\bf p}_1^*}{2E_1^*}\frac{d^3{\bf p}_2^*}{2E_2^*} \;
  \delta^{(3)}({\bf p}_1^*+{\bf p}_2^*) \delta(E^*-E_1^*-E_2^*)
  |\langle0|j_{\bf b}(0)|\pi\pi\rangle|^2 e^{-Et}
  \nonumber\\
  & = &
  \frac{1}{(2\pi)^2}\frac{2}{3} \int dE\; 
  g(\gamma)^2\frac{k^3}{E^*}|F_\pi(s)|^2 e^{-Et},
\end{eqnarray}
with $s=E^{*2}=4(m_\pi^2+k^2)$.

On the other hand, when taking a large volume limit in
(\ref{eq:corr_finite}), the summation over discrete energy states will
change to a continuum integral
\begin{equation}
  \sum_n\rightarrow \int dE\,\rho_V(E),
  \quad 
  \rho_V(E) = \frac{dn}{dE} =
  \frac{1}{\pi}\frac{d(\delta_1+\phi^{{\bf P},\Gamma})}{dE}
  = 
  \frac{E}{4\pi k^2} \left(
    k\frac{\partial \delta_1}{\partial k} +
    q\frac{\partial\phi^{{\bf P},\Gamma}}{\partial q}
  \right),
\end{equation}
where we have used the L\"uscher's quantization condition
(\ref{eq:Luscher_formula}). 
The correlator is now given by
\begin{equation}
  \label{eq:large_V2}
  C_V(t) \xrightarrow[V\rightarrow\infty]{} 
  \int dE\,\rho_V(E)|\langle0|j_{\bf b}|\pi\pi,n \rangle_V|^2 e^{-E_nt}.
\end{equation}

Comparing (\ref{eq:large_V2}) and (\ref{eq:large_V1}) we obtain the
relation (\ref{eq:LL_formula}).
Strictly speaking, the equivalent integral does not mean the
equivalent integrand. 
Also, in the demonstration we have used the L\"uscher's
quantization condition, which is only valid in the elastic scattering
region. 
However, the integrals given by (\ref{eq:large_V2}) and
(\ref{eq:large_V1}) cover also the inelastic scattering region.
To make a more rigorous demonstration, one can extend the approach of 
\cite{Meyer:2011um} to the moving frames by requiring that the $W$
particle  carry the nonzero momentum. 
This is very similar to the extension of the Lellouch-L\"uscher
formula \cite{Lellouch:2000pv} to the moving frames  \cite{Christ:2005gi,Kim:2005gf}.

\bibliography{time_like_FF}

\begin{thebibliography}{10}

\bibitem{Dudek:2006ut}
J.~J. Dudek and R.~G. Edwards,
\newblock Phys.Rev.Lett. {\bf 97}, 172001 (2006), hep-ph/0607140.

\bibitem{Cohen:2008ue}
S.~D. Cohen, H.-W. Lin, J.~Dudek, and R.~G. Edwards,
\newblock PoS {\bf LATTICE2008}, 159 (2008), 0810.5550.

\bibitem{Shintani:2011vc}
E.~Shintani, S.~Aoki, S.~Hashimoto, T.~Onogi, and N.~Yamada,
\newblock PoS {\bf LATTICE2010}, 159 (2010), 1102.5544.

\bibitem{Feng:2011za}
JLQCD Collaboration, X.~Feng {\em et~al.},
\newblock PoS {\bf LATTICE2011}, 154 (2011).

\bibitem{Feng:2012ck}
X.~Feng {\em et~al.},
\newblock Phys.Rev.Lett. {\bf 109}, 182001 (2012), 1206.1375.

\bibitem{Lin:2013im}
H.-W. Lin and S.~D. Cohen,
\newblock PoS {\bf ConfinementX}, 113 (2012), 1302.0874.

\bibitem{Blum:2011ng}
T.~Blum {\em et~al.},
\newblock Phys.Rev.Lett. {\bf 108}, 141601 (2012), 1111.1699.

\bibitem{Blum:2012uk}
T.~Blum {\em et~al.},
\newblock Phys.Rev. {\bf D86}, 074513 (2012), 1206.5142.

\bibitem{Boyle:2012ys}
RBC Collaboration, UKQCD Collaboration, P.~Boyle {\em et~al.},
\newblock Phys.Rev.Lett. {\bf 110}, 152001 (2013), 1212.1474.

\bibitem{Lellouch:2000pv}
L.~Lellouch and M.~Luscher,
\newblock Commun.Math.Phys. {\bf 219}, 31 (2001), hep-lat/0003023.

\bibitem{Aoki:2007rd}
CP-PACS Collaboration, S.~Aoki {\em et~al.},
\newblock Phys.Rev. {\bf D76}, 094506 (2007), 0708.3705.

\bibitem{Gockeler:2008kc}
QCDSF Collaboration, M.~Gockeler {\em et~al.},
\newblock PoS {\bf LATTICE2008}, 136 (2008), 0810.5337.

\bibitem{Feng:2010es}
X.~Feng, K.~Jansen, and D.~B. Renner,
\newblock Phys.Rev. {\bf D83}, 094505 (2011), 1011.5288.

\bibitem{Lang:2011mn}
C.~B. Lang, D.~Mohler, S.~Prelovsek, and M.~Vidmar,
\newblock Phys.Rev. {\bf D84}, 054503 (2011), 1105.5636.

\bibitem{Aoki:2011yj}
PACS-CS Collaboration, S.~Aoki {\em et~al.},
\newblock Phys.Rev. {\bf D84}, 094505 (2011), 1106.5365.

\bibitem{Pelissier:2012pi}
C.~Pelissier and A.~Alexandru,
\newblock Phys.Rev. {\bf D87}, 014503 (2013), 1211.0092.

\bibitem{Dudek:2012xn}
J.~J. Dudek, R.~G. Edwards, and C.~E. Thomas,
\newblock Phys.Rev. {\bf D87}, 034505 (2013), 1212.0830.

\bibitem{Brommel:2006ww}
QCDSF/UKQCD Collaboration, D.~Brommel {\em et~al.},
\newblock Eur.Phys.J. {\bf C51}, 335 (2007), hep-lat/0608021.

\bibitem{Frezzotti:2008dr}
ETM Collaboration, R.~Frezzotti, V.~Lubicz, and S.~Simula,
\newblock Phys.Rev. {\bf D79}, 074506 (2009), 0812.4042.

\bibitem{Boyle:2008yd}
P.~Boyle {\em et~al.},
\newblock JHEP {\bf 0807}, 112 (2008), 0804.3971.

\bibitem{Aoki:2009qn}
JLQCD Collaboration, TWQCD Collaboration, S.~Aoki {\em et~al.},
\newblock Phys.Rev. {\bf D80}, 034508 (2009), 0905.2465.

\bibitem{Nguyen:2011ek}
O.~H. Nguyen, K.-I. Ishikawa, A.~Ukawa, and N.~Ukita,
\newblock JHEP {\bf 1104}, 122 (2011), 1102.3652.

\bibitem{Brandt:2013dua}
B.~B. Brandt, A.~Juttner, and H.~Wittig,
\newblock (2013), 1306.2916.

\bibitem{Koponen:2013boa}
J.~Koponen, F.~Bursa, C.~Davies, G.~Donald, and R.~Dowdall,
\newblock PoS {\bf LATTICE2013}, 282 (2014), 1311.3513.

\bibitem{Christ:2005gi}
N.~H. Christ, C.~Kim, and T.~Yamazaki,
\newblock Phys.Rev. {\bf D72}, 114506 (2005), hep-lat/0507009.

\bibitem{Kim:2005gf}
C.~Kim, C.~Sachrajda, and S.~R. Sharpe,
\newblock Nucl.Phys. {\bf B727}, 218 (2005), hep-lat/0507006.

\bibitem{Meyer:2011um}
H.~B. Meyer,
\newblock Phys.Rev.Lett. {\bf 107}, 072002 (2011), 1105.1892.

\bibitem{Gounaris:1968mw}
G.~Gounaris and J.~Sakurai,
\newblock Phys.Rev.Lett. {\bf 21}, 244 (1968).

\bibitem{Protopopescu:1973sh}
S.~Protopopescu {\em et~al.},
\newblock Phys.Rev. {\bf D7}, 1279 (1973).

\bibitem{Estabrooks:1974vu}
P.~Estabrooks and A.~D. Martin,
\newblock Nucl.Phys. {\bf B79}, 301 (1974).

\bibitem{Akhmetshin:2006wh}
R.~Akhmetshin {\em et~al.},
\newblock JETP Lett. {\bf 84}, 413 (2006), hep-ex/0610016.

\bibitem{Akhmetshin:2006bx}
CMD-2 Collaboration, R.~Akhmetshin {\em et~al.},
\newblock Phys.Lett. {\bf B648}, 28 (2007), hep-ex/0610021.

\bibitem{Achasov:2006vp}
M.~Achasov {\em et~al.},
\newblock J.Exp.Theor.Phys. {\bf 103}, 380 (2006), hep-ex/0605013.

\bibitem{Ambrosino:2010bv}
KLOE Collaboration, F.~Ambrosino {\em et~al.},
\newblock Phys.Lett. {\bf B700}, 102 (2011), 1006.5313.

\bibitem{Beringer:1900zz}
Particle Data Group, J.~Beringer {\em et~al.},
\newblock Phys.Rev. {\bf D86}, 010001 (2012).

\bibitem{Jegerlehner:2011ti}
F.~Jegerlehner and R.~Szafron,
\newblock Eur.Phys.J. {\bf C71}, 1632 (2011), 1101.2872.

\bibitem{Schael:2005am}
ALEPH Collaboration, S.~Schael {\em et~al.},
\newblock Phys.Rept. {\bf 421}, 191 (2005), hep-ex/0506072.

\bibitem{Davier:2005xq}
M.~Davier, A.~Hocker, and Z.~Zhang,
\newblock Rev.Mod.Phys. {\bf 78}, 1043 (2006), hep-ph/0507078.

\bibitem{Melikhov:2002nf}
D.~Melikhov, O.~Nachtmann, and T.~Paulus,
\newblock (2002), hep-ph/0209151.

\bibitem{Bruch:2004py}
C.~Bruch, A.~Khodjamirian, and J.~H. Kuhn,
\newblock Eur.Phys.J. {\bf C39}, 41 (2005), hep-ph/0409080.

\bibitem{Luscher:1990ux}
M.~Luscher,
\newblock Nucl.Phys. {\bf B354}, 531 (1991).

\bibitem{Rummukainen:1995vs}
K.~Rummukainen and S.~A. Gottlieb,
\newblock Nucl.Phys. {\bf B450}, 397 (1995), hep-lat/9503028.

\bibitem{Lin:2001ek}
C.~D. Lin, G.~Martinelli, C.~T. Sachrajda, and M.~Testa,
\newblock Nucl.Phys. {\bf B619}, 467 (2001), hep-lat/0104006.

\bibitem{Aoki:2012pma}
S.~Aoki {\em et~al.},
\newblock PTEP {\bf 2012}, 01A106 (2012).

\bibitem{Fukaya:2006vs}
JLQCD Collaboration, H.~Fukaya {\em et~al.},
\newblock Phys.Rev. {\bf D74}, 094505 (2006), hep-lat/0607020.

\bibitem{Aoki:2007ka}
S.~Aoki, H.~Fukaya, S.~Hashimoto, and T.~Onogi,
\newblock Phys.Rev. {\bf D76}, 054508 (2007), 0707.0396.

\bibitem{Noaki:2009xi}
J.~Noaki {\em et~al.},
\newblock Phys.Rev. {\bf D81}, 034502 (2010), 0907.2751.

\bibitem{Foley:2005ac}
J.~Foley {\em et~al.},
\newblock Comput.Phys.Commun. {\bf 172}, 145 (2005), hep-lat/0505023.

\bibitem{Liu:2011jp}
Q.~Liu,
\newblock PoS {\bf LATTICE2011}, 287 (2011), 1110.2143.

\bibitem{Feng:2011ah}
ETM Collaboration, X.~Feng, K.~Jansen, and D.~B. Renner,
\newblock PoS {\bf LATTICE2010}, 104 (2010), 1104.0058.

\bibitem{Luscher:1990ck}
M.~Luscher and U.~Wolff,
\newblock Nucl.Phys. {\bf B339}, 222 (1990).

\bibitem{Bali:2005fu}
SESAM Collaboration, G.~S. Bali, H.~Neff, T.~Dussel, T.~Lippert, and
  K.~Schilling,
\newblock Phys.Rev. {\bf D71}, 114513 (2005), hep-lat/0505012.

\bibitem{Feng:2009ij}
X.~Feng, K.~Jansen, and D.~B. Renner,
\newblock Phys.Lett. {\bf B684}, 268 (2010), 0909.3255.

\bibitem{Blum:2011pu}
T.~Blum {\em et~al.},
\newblock Phys.Rev. {\bf D84}, 114503 (2011), 1106.2714.

\bibitem{Dudek:2012gj}
J.~J. Dudek, R.~G. Edwards, and C.~E. Thomas,
\newblock Phys.Rev. {\bf D86}, 034031 (2012), 1203.6041.

\bibitem{Kaneko:2010ru}
JLQCD Collaboration, T.~Kaneko {\em et~al.},
\newblock PoS {\bf LATTICE2010}, 146 (2010), 1012.0137.

\bibitem{Fukaya:2014jka}
H.~Fukaya {\em et~al.},
\newblock Phys.Rev. {\bf D90}, 034506 (2014), 1405.4077.

\bibitem{Liu:2008hy}
G.~Meng and C.~Liu,
\newblock Phys.Rev. {\bf D78}, 074506 (2008), 0804.4308.

\bibitem{Niu:2009gt}
Z.-Y. Niu, M.~Gong, C.~Liu, and Y.~Shen,
\newblock Phys.Rev. {\bf D80}, 114509 (2009), 0909.3154.

\bibitem{Niu:2010rk}
Z.-Y. Niu {\em et~al.},
\newblock Phys.Rev. {\bf D82}, 054501 (2010), 1005.5571.

\bibitem{Francis:2013fzp}
A.~Francis, B.~Jager, H.~B. Meyer, and H.~Wittig,
\newblock Phys.Rev. {\bf D88}, 054502 (2013), 1306.2532.

\bibitem{Luscher:1991cf}
M.~Luscher,
\newblock Nucl.Phys. {\bf B364}, 237 (1991).

\bibitem{Leskovec:2012gb}
L.~Leskovec and S.~Prelovsek,
\newblock Phys.Rev. {\bf D85}, 114507 (2012), 1202.2145.

\end{thebibliography}
\bibliographystyle{h-physrev}

\end{document}